\documentclass[prd,showpcs,amsmath,amssymb,nofootinbib,longbibliography,twocolumn,showpacs,notitlepage]{revtex4-1}
\usepackage{multirow}
\usepackage{epsfig}
\usepackage{amsmath}
\usepackage{bm}
\usepackage{times}
\usepackage{graphicx}
\usepackage{color}
\usepackage{slashed}
\usepackage{graphicx}
\usepackage{amsmath}
\usepackage[latin1]{inputenc}
\usepackage{hyperref}
\usepackage{soul}
\usepackage{lipsum}

\def\bea{\begin{eqnarray}}
\def\eea{\end{eqnarray}}
\def\bean{\begin{equation*}}
\def\eean{\end{equation*}} 
\def\nn{\nonumber}
\def\beaal{\begin{align}}
\def\eeaal{\end{align}}

\begin{document}

\title{Baryon and Lepton Number Violation from Gravitational Waves}

\author{Bartosz~Fornal}
\affiliation{\vspace{1mm}Department of Physics and Astronomy, University of Utah, Salt Lake City, UT 84112, USA\vspace{1mm}}
\author{Barmak~Shams~Es~Haghi}
\affiliation{\vspace{1mm}Department of Physics and Astronomy, University of Utah, Salt Lake City, UT 84112, USA\vspace{1mm}}
\date{\today}
\begin{abstract}
\vspace{1mm}
We describe a unique gravitational wave signature for a class of models with a vast hierarchy between the symmetry breaking scales.
The unusual shape of the signal is a result of the overlapping contributions to the stochastic gravitational wave background from cosmic strings produced at a high scale and a cosmological phase transition at a low scale. 
We apply this idea to a simple model with gauged baryon and lepton number, in which the high-scale breaking of lepton number is motivated by the seesaw mechanism for the neutrinos, whereas the low scale of baryon number breaking is required by the observed dark matter relic density. 
The novel signature can be searched for in upcoming gravitational wave experiments. 
\vspace{10mm}
\end{abstract}

\maketitle

\section{Introduction}

Gravitational wave detectors opened the door to an entirely new set of opportunities for probing unexplored avenues in physics and astronomy. 
Many astrophysical discoveries have already been made using solely  the data gathered  by LIGO \cite{TheLIGOScientific:2014jea} and Virgo \cite{TheVirgo:2014hva} in their initial runs. One can only imagine what will be learned from future detectors like  LISA \cite{Audley:2017drz}, DECIGO \cite{Kawamura:2011zz}, Cosmic Explorer \cite{Reitze:2019iox}, Einstein Telescope \cite{Reitze:2019iox} and Big Bang Observer \cite{Crowder:2005nr}.
Interestingly, gravitational wave experiments may  not only reveal  information about black hole and neutron star mergers, but they can also provide a deep insight into the particle physics of the early universe.

The shape of the stochastic gravitational wave background is the key to unraveling  the symmetry breaking  pattern in the first instances after the Big Bang. 
It enables us  to explore the physics at the very high energy scale,  inaccessible directly in any other experiment and, so far, probed only indirectly, e.g., via proton decay searches. 
Thus, a discovery of a gravitational wave signal from the early universe  would provide invaluable  insight into the UV completion of the Standard Model. 
The two  main sources of such gravitational waves are  cosmological phase transitions and cosmic strings. 

Depending on the parameters of the scalar potential, upon symmetry breaking at early times the universe might have been trapped in a vacuum which became metastable as the temperature dropped. In the presence of a potential barrier separating this false vacuum from the true vacuum, a first order phase transition would occur, nucleating bubbles of true vacuum which expanded and populated the universe. This, in turn, would lead to a production of  gravitational waves with a characteristic bump-like shape in the spectrum.  Indeed, such signals have been analyzed in the context of various particle physics models (see, e.g., \cite{Apreda:2001us,Grojean:2006bp,Leitao:2012tx,Schwaller:2015tja,Huang:2017laj,Huang:2017rzf,Demidov:2017lzf,Hashino:2018zsi,Okada:2018xdh,Ahriche:2018rao,Brdar:2018num,Croon:2018kqn,Angelescu:2018dkk,Hasegawa:2019amx,Dev:2019njv,Brdar:2019fur,Wang:2019pet,Greljo:2019xan,vonHarling:2019gme,Hall:2019ank,Huang:2020bbe,Fornal:2020ngq}).

On the other hand, symmetry breaking can also lead  to the production topological defects such as cosmic strings, as it happens, e.g., in models with a complex scalar field charged under a ${\rm U}(1)$ gauge group. A network of cosmic strings is a long-lasting source of gravitational waves and gives rise to a spectrum which, to a good approximation, is flat in a wide range of frequencies. Such cosmic string signatures have also been considered  in many models (see, e.g., \cite{Blanco-Pillado:2017oxo,Ringeval:2017eww,Cui:2017ufi,Cui:2018rwi,Guedes:2018afo,Dror:2019syi,Gouttenoire:2019kij,Gouttenoire:2019rtn,Buchmuller:2019gfy,King:2020hyd,Zhou:2020ils}).

For a single broken ${\rm U}(1)$ gauge group, the  cosmic string contribution is negligible in the frequency range relevant for the gravitational wave signal of a phase transition.  However,  if cosmic strings were produced by a high-scale breaking of ${\rm U}(1)_{\rm high}$, whereas the first order phase transition was triggered by the breaking of a different ${\rm U}(1)_{\rm low}$ at a much lower energy scale,  the two contributions could end up comparable at frequencies corresponding to the ${\rm U}(1)_{\rm low}$ breaking.

In this paper, we propose to look precisely for such a combined signature of a phase transition and cosmic strings. This is naturally realized in models with a seesaw mechanism, in which the large mass of the right-handed neutrinos arises from  lepton number breaking at a scale $v_L\sim 10^{10}-10^{15} \ {\rm GeV}$.  In a certain frequency band, the resulting cosmic string signal  can have a similar magnitude to that of a phase transition happening at a scale $v_B\sim 10^3 - 10^5 \ {\rm GeV}$.
For a particular realization of this scenario, we focus on a simple model with gauged baryon and lepton number  \cite{Duerr:2013dza}, where the high-scale lepton number breaking is motivated by the seesaw mechanism, whereas the low scale of baryon number breaking is required to explain the dark matter relic abundance.

There are several reasons for gauging baryon and lepton number, and breaking them spontaneously. 
In the Standard Model, $B$ and $L$ are  accidental global symmetries.  Such symmetries, however, cannot be fundamental  in a consistent theory of quantum gravity, unlike gauge symmetries \cite{Banks:2010zn}.\break Another motivation  comes from the matter-antimatter asymmetry of the universe, which  requires baryon number to be broken beyond the nonperturbative effects mediated by the electroweak sphalerons. 
Early attempts of gauging ${\rm U}(1)_B$ and  ${\rm U}(1)_L$ were made in \cite{Pais:1973mi,Rajpoot:1987yg,Foot:1989ts,Carone:1995pu,Georgi:1996ei}, but phenomenologically viable models consistent with LHC constraints were constructed only recently \cite{FileviezPerez:2010gw,Duerr:2013dza,Schwaller:2013hqa,Perez:2014qfa,Duerr:2014wra,Arnold:2013qja}, and can  naturally account for the  non-observation of proton decay, small neutrino masses, dark\break matter and the matter-antimatter asymmetry. 
This  idea was further generalized to the non-Abelian case and partially unified theories in \cite{Fornal:2015boa,Fornal:2015one,FileviezPerez:2016laj,Fornal:2017owa}.

Although the focus of this paper is on the model constructed in \cite{Duerr:2013dza}, our proposed signature  is much more general and can be realized in other models with two ${\rm U}(1)$ gauge groups broken at vastly different scales, not necessarily associated with baryon or lepton number.

\section{Gauging baryon and lepton number}\label{sec2}
The model we consider is based on the gauge group
\bea
{\rm SU}(3)_c \times {\rm SU}(2)_L \times {\rm U}(1)_Y \times {\rm U}(1)_B \times {\rm U}(1)_L \ .
\eea
The charges under ${\rm U}(1)_B$ and ${\rm U}(1)_L$ of the Standard Model particles are the same as their standard $B$ and $L$ assignments. 
The requirement of gauge anomaly cancellation implies the existence of new fermion fields.  There are many possible choices for these extra fields, with various  hypercharge and ${\rm U}(1)_B$ and ${\rm U}(1)_L$  charge assignments. Since the particular choice of the new fermions or their charges does not  qualitatively impact our results, we consider the original set of fields proposed in \cite{Duerr:2013dza}.

The Standard Model is extended with  three families of  right-handed neutrinos $\nu_{iR}$  and  the following single set of\break leptobaryonic fermions,
\bea
\Psi_L& =& \left(1,2,\tfrac12,B_1,L_1\right)  ,  \ \ \ \, \Psi_R = \left(1,2,\tfrac12,B_2,L_2\right)  , \nn\\
\eta_R\,& =& \left(1,1,1,B_1,L_1\right) , \, \ \ \ \ \,  \eta_L\, = \left(1,1,1,B_2,L_2\right)  , \nn\\
\chi_R& =& \left(1,1,0,B_1,L_1\right) , \, \ \ \ \ \,\chi_L = \left(1,1,0,B_2,L_2\right) , \ \ \ \ \  \ 
\eea
with the relations $B_2-B_1 = 3$ and $L_2-L_1 = 3$ satisfied.
We are going to assume $B_1=L_1=-1$ and $B_2=L_2=2$. \vspace{1mm}

\noindent
Two new scalar fields are also introduced into the model,
\bea
\Phi_L = (1,1,0,0,-2)   \ , \ \ \ \ \Phi_B = (1,1,0,-3,-3)\ . \ \ \ \ 
\eea
The gauge group ${\rm U}(1)_L$ is broken by  the  vacuum expectation value (vev) of the field $\Phi_L$ at a high scale, whereas 
${\rm U}(1)_B$ is broken by the vev of $\Phi_B$ at a lower scale,
\bea
\langle \Phi_L \rangle = \frac{v_L}{\sqrt2} \ , \ \ \ \ \ \ \langle \Phi_B \rangle = \frac{v_B}{\sqrt2} \ ,
\eea 
with a  large hierarchy between them, and with respect to the Standard Model Higgs $H$ vev $v\approx 246 \ {\rm GeV}$,
\bea\label{hier}
v_L \gg v_B \gg v \ .
\eea

The Lagrangian terms describing the interactions of the new fermions with  scalars are,
\bea
\mathcal{L} &\,\supset\,& Y_{\Psi} \,\overline{\Psi}_L \Psi_R \Phi_B + Y_{\eta}\, \overline{\eta}_R \eta_L \Phi_B + Y_{\chi}\, \overline{\chi}_R \chi_L \Phi_B\nn\\
&+&y_{\eta_R} \overline{\Psi}_L H \eta_R +  y_{\eta_L} \overline{\Psi}_R H \eta_L + y_{\chi_R} \overline{\Psi}_L H \chi_R \nn\\
&+&  y_{\chi_L} \overline{\Psi}_R H \chi_L + y_\nu \bar{l}_L \tilde{H}\nu_R  + Y_\nu \nu_R \nu_R \Phi_L  + {\rm h.c.}\ . \ \ \ \ \ 
\eea
The only fermions which couple to $\Phi_L$ are the right-handed neutrinos. Due to the large hierarchy between the symmetry breaking scales, one can ignore the effect of  terms involving $H$ on the fermion masses. \vspace{1mm}

The scalar potential for $\Phi_L$ and $\Phi_B$ is given by
\bea
&&V(\Phi_L, \Phi_B)\nn\\
&&= -\mu_L^2 |\Phi_L|^2 + \lambda_L |\Phi_L|^4 -\mu_B^2 |\Phi_B|^2 + \lambda_B |\Phi_B|^4 \ , \ \ \ 
\eea
where we have left out the possible cross terms  between the fields $\Phi_L$, $\Phi_B$ and $H$, assuming that the corresponding coefficients are small.

The Standard Model covariant derivative is extended to 
\bea
D_\mu = D_\mu^{\rm SM} + i g_B B_\mu B + i g_L L_\mu L \ .
\eea
After symmetry breaking, the fields $B_\mu$ and $L_\mu$ give rise to the gauge bosons $Z_B$ and $Z_L$ that couple exclusively to baryons and leptons, respectively. Their masses are
\bea
m_{Z_B} = 3\, g_Bv_B \ , \ \ \ \ \ m_{Z_L} = 2 \,g_Lv_L \ .
\eea
 Since ${\rm U}(1)_B$ and ${\rm U}(1)_L$ are not unified with the Standard Model gauge group, the gauge couplings $g_B$ and  $g_L$ are free parameters, similarly to $\lambda_B$, $\lambda_L$ and the new Yukawas.

The breaking of ${\rm U}(1)_L$ results in a $\Delta L=2$ mass term for the right-handed neutrinos, which leads to  the type I  seesaw mechanism.
Assuming the Yukawa couplings  $y_\nu \sim \mathcal{O}(10^{-2})$ and $Y_\nu \sim \mathcal{O}(1)$, the measured neutrino mass splittings are naturally explained if  the scale of  lepton number  breaking is 
\bea
v_L \approx 10^{11} \ {\rm GeV} \ .
\eea

The breaking of ${\rm U}(1)_B$ leads to baryon number violation, but only by three units, $\Delta B =3$. This implies that proton is absolutely stable in this model. In addition, since the lowest-dimensional baryon number violating operator  appears at\break dimension  nineteen,
\bea
{\mathcal{O}} \sim \frac{(u_R u_R d_R e_R)^3 \,\Phi_B}{\Lambda^{15}} \ ,
\eea
the resulting $\Delta B = 3$ processes are highly suppressed.

Interestingly, after  ${\rm U}(1)_B$ and ${\rm U}(1)_L$ breaking, an accidental global symmetry remains  in the new sector, under which the leptobaryons transform as
\bea
&&\Psi_{L,R} \to e^{i\alpha}\,\Psi_{L,R} \ , \ \ \ \eta_{L,R} \to e^{i\alpha}\,\eta_{L,R} \ , \nn\\
&&\,\chi_{L,R} \to e^{i\alpha}\,\chi_{L,R}  \ .
\eea
This implies the stability of the lightest leptobaryon. Assuming $Y_\chi < Y_{\Psi, \eta}$,  the lightest new field is  $\chi$. As a Standard Model singlet, $\chi$ becomes a viable dark matter candidate.

If  $\chi$ is indeed a  dark matter particle,  the cross section for its annihilation needs to be consistent with the 
observed dark matter relic density  $h^2 \Omega_{\rm DM} =0.12$ \cite{Aghanim:2018eyx}.
Since the  annihilation proceeds via the $s$-channel $\chi\,\bar{\chi} \to Z_B^* \to q\,\bar{q}$, and
the mass of the gauge boson $Z_B$ depends on the symmetry breaking scale, this introduces a non-trivial relation between the parameters $v_B$, $g_B$ and $Y_\chi$. As pointed out in 
\cite{Duerr:2014wra}, with $B_1+B_2 = 1$ this imposes the upper bound
\bea\label{bbb}
g_Bv_B \lesssim 20 \ {\rm TeV} \ .
\eea
In the analysis of the phase transition resulting from ${\rm U}(1)_B$ breaking in Sec.\,\ref{ptt} we adopt $\lambda_B \sim 10^{-2}$. The gravitational wave signal is then maximized for $g_B \sim 0.3$. Therefore, as a benchmark scenario we assume the parameter values,
\bea\label{bench}
v_B=20 \ {\rm TeV} ,  \ \, g_B = 0.3  \,,  \ \, \lambda_B = 10^{-2}  , \  \, Y_\chi = 0.6 \,, \ \ \ \ \ \ 
\eea
with $Y_{\Psi,\eta}$ slightly larger than $Y_\chi$, so that $\chi$ remains the lightest leptobaryon. In this benchmark model $M_{Z_B} \sim 20 \ {\rm TeV}$, well beyond the LHC reach. However, a phase transition signal associated with symmetry breaking at this scale is within the reach of upcoming gravitational wave experiments.

\section{Cosmic strings from ${\bf{{U}(1)_{\textit{L}}}}$ breaking}

%We first determine  the gravitational wave signal from the breaking of gauged lepton number. With  $y_\nu, Y_\nu \sim \mathcal{O}(1)$, the  seesaw mechanism points to $v_L \sim 10^{15} \ {\rm GeV}$. In general,\break however, the Yukawas can be much smaller. In the Standard Model itself the charged leptons have $y_l \sim 10^{-6} - 10^{-2}$. Assuming $Y_\nu \sim \mathcal{O}(1)$ and $y_\nu \sim \mathcal{O}(10^{-6})$, the lepton number breaking scale becomes $v_L \sim 10^{9} \ {\rm GeV}$. To account for the freedom in the choice of Yukawas, we analyze the signature of ${\rm U}(1)_L$  breaking for  $v_L \sim 10^{10} - 10^{13} \ {\rm GeV}$. 

The spontaneously broken gauge symmetry ${\rm U}(1)_L$ can lead to gravitational wave production  in two very different ways: $(1)$ within a very short timescale  during a phase transition via sound waves, bubble collisions and turbulence (as discussed in Sec.\,\ref{ptt}), or $(2)$ in a long-term process resulting from the dynamics of the produced cosmic strings.  Since the  frequencies of the type $(1)$ signal for a high scale $v_L$ are inaccessible via gravitational detectors in the foreseeable future, we consider only the cosmic string signature of  ${\rm U}(1)_L$ breaking, since it extends to lower frequencies.

\subsection{Cosmic string network}

Cosmic strings may be  generated during  the   breaking  of ${\rm U}(1)_L$ through the Kibble mechanism~\cite{Kibble:1976sj}. They are topological defects corresponding to one-dimensional field configurations where the symmetry remains unbroken. The produced cosmic string network is  characterized by the string tension $\mu$ (energy per unit length). It is related to the symmetry breaking scale $v_L$  via \cite{Vilenkin:2000jqa,Gouttenoire:2019kij}
\bea
G\mu =2\pi \left(\frac{v_L}{M_P}\right)^2 ,
\eea
where $G$ is the gravitational constant, $M_P=1.22\times 10^{19} \ {\rm GeV}$ is the Planck mass, and we assumed that the winding number $n=1$.
Measurements of the cosmic microwave background constrain the string tension to be $G\mu\lesssim 10^{-7}$ \cite{Ade:2013xla}.

The cosmic string network experiences two competing contributions to its dynamics: stretching (due to the expansion of the universe) and formation of string loops (when long strings intersect and intercommute). The string loops themselves are unstable: they oscillate and eventually decay. 
The combination of the two effects results in a \textit{scaling regime} that consists of a small number of Hubble-length  strings and a large number of string loops \cite{Kibble:1984hp, PhysRevLett.60.257, PhysRevLett.63.2776, PhysRevD.40.973, PhysRevLett.64.119}. 
The energy is continuously transferred from long strings to loops, and eventually to radiation or particles, constituting a fixed fraction of the total energy density of the universe \cite{Hindmarsh:1994re}. 

The dominant decay channel of string loops is gravitational radiation \cite{Olum:1999sg, Moore:2001px}. 
In particular, powerful bursts of gravitational waves are expected to be produced by cusps and kinks propagating along the string loops, as well as  kink-kink collisions. 
 The superposition of these bursts results in  a stochastic gravitational wave 
background. 
To calculate the corresponding  signal, we follow the framework adopted in \cite{Cui:2018rwi,Gouttenoire:2019kij}.

\subsection{String dynamics}

Let us consider a cosmic string loop created at time $t_i$  and of length $l(t_i)=\alpha\,t_i$, where $\alpha$ is an approximately constant loop size parameter.  We assume $\alpha=0.1$, as this provides a good approximation for the loop size distribution determined in \cite{Blanco-Pillado:2013qja,Blanco-Pillado:2017oxo}.
While the loop oscillates, it emits gravitational waves with frequencies 
\bea
\tilde\nu = \frac{2k}{l}  \ ,  \ \ \  \ {\rm where} \ \ \ k \in \mathbb{Z}^+ \ .
\eea
In the current epoch, this corresponds to   $\nu = a({\tilde{t}})/a(t_0) \,\tilde\nu$, where $a$ is the scale factor of the universe,  $\tilde{t}$ is the  emission time and  $t_0$ denotes the time today.
\vspace{1mm}

The spectrum of  gravitational waves emitted from a single loop is given by  \cite{Blanco-Pillado:2013qja,Blanco-Pillado:2017oxo}
\bea
P_{\rm CS}^{(k,n)} = \frac{\Gamma \,G\mu^2 \, k^{-n}}{\sum_{p=1}^\infty {p}^{-n}} \ ,
\eea
where $n=\frac43$, $\frac53$, $2$ corresponds to the contribution from cusps, kinks and kink-kink collisions, respectively, and $\Gamma \simeq 50$ \cite{PhysRevD.31.3052}.
Due to the emission of gravitational waves, the loop shrinks,
\bea
l(\tilde{t}) = \alpha\,t_i - \Gamma \,G\mu \,(\tilde{t}-t_i) \ ,
\eea
and decays after the time $\,\tau = {\alpha\,t_i}/{(\Gamma G \mu)}$. 
Furthermore, it was shown in \cite{Blanco-Pillado:2013qja} that only $\mathcal{F}_\alpha \approx 10\%$ of the loops contribute to the gravitational wave signal. %as the remaining majority of the energy is lost by long strings going into highly boosted smaller loops which provide only a  subdominant contribution. 
\vspace{1mm}

The model-dependence enters through the assumption regarding the loop distribution function $f(l,t)\,dl$, which describes the number density of loops with an invariant length in the range $(l, l+dl)$ at the cosmic time $t$. 
We adopt the framework of the Velocity-Dependent One-Scale model \cite{Martins:1995tg,Martins:1996jp,Martins:2000cs},\break which describes the evolution
of a string network  in terms of only two parameters: the mean string velocity and the correlation length. In the scaling solution regime, for which those parameters become constant, one arrives at
\bea
f(l,t_i) = \frac{\sqrt2 \,{C}_{\rm eff}}{\alpha\,t_i^4}\,\delta(l-\alpha\,t_i) \ ,
\eea
where ${C}_{\rm eff}=5.4$ for the radiation era and ${C}_{\rm eff}=0.39$ for matter domination \cite{Cui:2018rwi}.

\begin{figure}[t!]
\includegraphics[trim={1.3cm 0.5cm 1.2cm 0},clip,width=9.0cm]{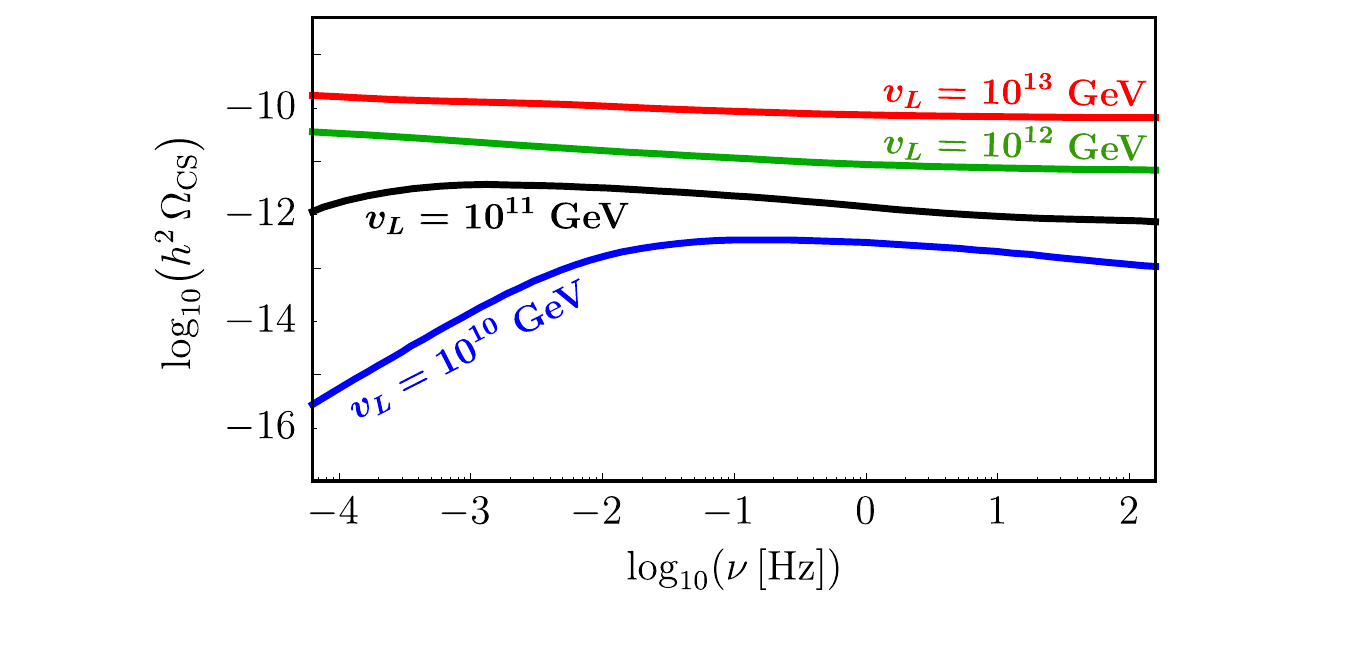} \vspace{-5mm}
\caption{Stochastic gravitational wave  background from the cosmic string network produced from gauged ${\rm U}(1)_L$ breaking for several choices of the scale $v_L$.\\}\label{fCS}
\end{figure}

\subsection{Gravitational wave signal}

The dynamics of the cosmic string network generates  the stochastic gravitational wave background  given  by  \cite{Cui:2018rwi,Gouttenoire:2019kij}
\bea\label{CSspec}
h^2\Omega_{\rm CS}(\nu) &=&\frac{2 h^2\mathcal{F}_\alpha}{\rho_c\,\nu\,\alpha^2}\,\sum_{k,n}\,{k\, P_{\rm CS}^{(k,n)}} \int_{t_F}^{t_0}\!d\tilde t\ \frac{C_\text{eff}(t_{i,k})}{t_{i,k}^{\, 4}}\ \ \ \ \nn\\
&\times&\left(\frac{a(\tilde t)}{a(t_0)}\right)^{\!5}\left(\frac{a(t_{i,k})}{a(\tilde t)}\right)^{\!3}\Theta(t_{i,k}-t_{F}) \ .
\eea
In the expression above $\rho_c$ is the critical density, $\Theta$ is the Heaviside function, $t_F$ is the time of the cosmic string network formation (i.e., when the energy scale of the\break universe is equal to the string tension, $\sqrt{\rho(t_F)}=\mu$ \cite{Gouttenoire:2019kij}),\break $t_{i,k}$ is the time of the loop production,
\bea
t_{i,k}\equiv t_{i,k}(\tilde t,\nu)=\frac{1}{\alpha}\left(\frac{2k}{\nu}\frac{a(\tilde t)}{a(t_0)}+\Gamma \,G\mu \,\tilde t\right),
\eea
and, as before, $\tilde{t}$ is the time of the gravitational wave emission and $t_0$ is the time today. 
We have also assumed  $G\mu \ll \alpha$. Following \cite{Cui:2018rwi,Gouttenoire:2019kij}, we consider only the contribution of the cusps to the gravitational wave signal.
\vspace{1mm}

Figure \ref{fCS} shows the resulting cosmic string contribution $h^2\Omega_{\rm CS}$ to the stochastic gravitational wave background, calculated using   Eq.\,(\ref{CSspec}), for several choices of the ${\rm U}(1)_L$ breaking scale, and in the frequency range relevant for  upcoming gravitational wave experiments.

\section{Phase transition from ${\bf{{U}(1)_{\textit{B}}}}$ breaking}\label{ptt}

In this section we derive the spectrum of the stochastic gravitational wave background generated by sound waves, bubble collisions and magnetohydrodynamic turbulence  from the first order phase transition triggered by gauged baryon number breaking. 
For the sound wave contribution we adopt a novel estimate of the  suppression factor recently derived in \cite{Guo:2020grp}. This weakens the signal from  sound waves  to such an extent that the effect of bubble wall collisions becomes dominant  at lower frequencies, which is typically the case only at higher frequencies.

%The scale of ${\rm U}(1)_B$ breaking in the model cannot be high due to the dark matter relic density constraints, as explained in Sec.\ref{sec2}. Given the fact that the LHC searches for $Z_B$ impose also a lower bound on the symmetry breaking scale of $\sim{\rm TeV}$, the benchmark scenario  in Eq.\,(\ref{bench}) provides a qualitatively good representation of the allowed parameter space.

\subsection{Effective potential}

The large hierarchy between the scales, as given by Eq.\,(\ref{hier}), implies that the effective potential for the background field $\phi_B \equiv {\sqrt2}\,{\rm Re}(\Phi_B)$ can be considered independently from the other background fields.  The three types of contributions to the potential are: tree-level,  one-loop  and finite temperature. \vspace{1mm}

The tree level part  is 
\bea
V_{\rm tree}(\phi_B) = -\frac12 \lambda_B v_B^2 \phi_B^2 + \frac14 \lambda_B \phi_B^4 \ ,
\eea
where we  used the relation between the parameters satisfied at the minimum, $\mu_B = v_B\sqrt{\lambda_B}$. 

The one-loop Coleman-Weinberg contribution, adopting the cutoff regularization scheme and matching the one-loop and tree-level minima, can be written as
\bea\label{1l}
V_{\rm 1\text{-}loop}(\phi_B) &=& \sum_{i} \frac{n_i}{32\pi^2}\bigg\{m_i^4(\phi_B)\left[\log\left(\frac{m_i(\phi_B)}{m_i(v_B)}\right)-\frac34\right]\nn\\
&&+\  m_i^2(\phi_B)\, m_i^2(v_B)\bigg\} \ ,
\eea
where the sum is over all  particles  coupling to $\phi_B$, while $n_i$ is the number of degrees of freedom of a given particle, with a minus sign for fermions. For the Goldstone boson $\chi_B$,  one needs to replace  $m_{\chi_B}(v_B) \to m_{\phi_B}(v_B)$. 
The background field-dependent masses are
\bea
&& m_{Z_B}(\phi_B) = 3\,g_B \phi_B  \ , \ \ \ \ m_{\phi_B}(\phi_B) = [\lambda_B ({3\phi_B^2 - v_B^2})]^{1/2}  \ , \nn\\
&&m_{\chi_B}(\phi_B) = [\lambda_B ({\phi_B^2 - v_B^2})]^{1/2}  , \ \ \ \ m_\Psi (\phi_B) = Y_\Psi \phi_B/\sqrt2 \ , \nn\\
&&m_\eta (\phi_B) = Y_\eta \phi_B/\sqrt2  \ , \ \ \ \ \ m_\chi (\phi_B) = Y_\chi \phi_B/\sqrt2  \ . 
\eea

\begin{figure}[t!]
\includegraphics[trim={1.0cm 0.6cm 1.2cm 0},clip,width=8cm]{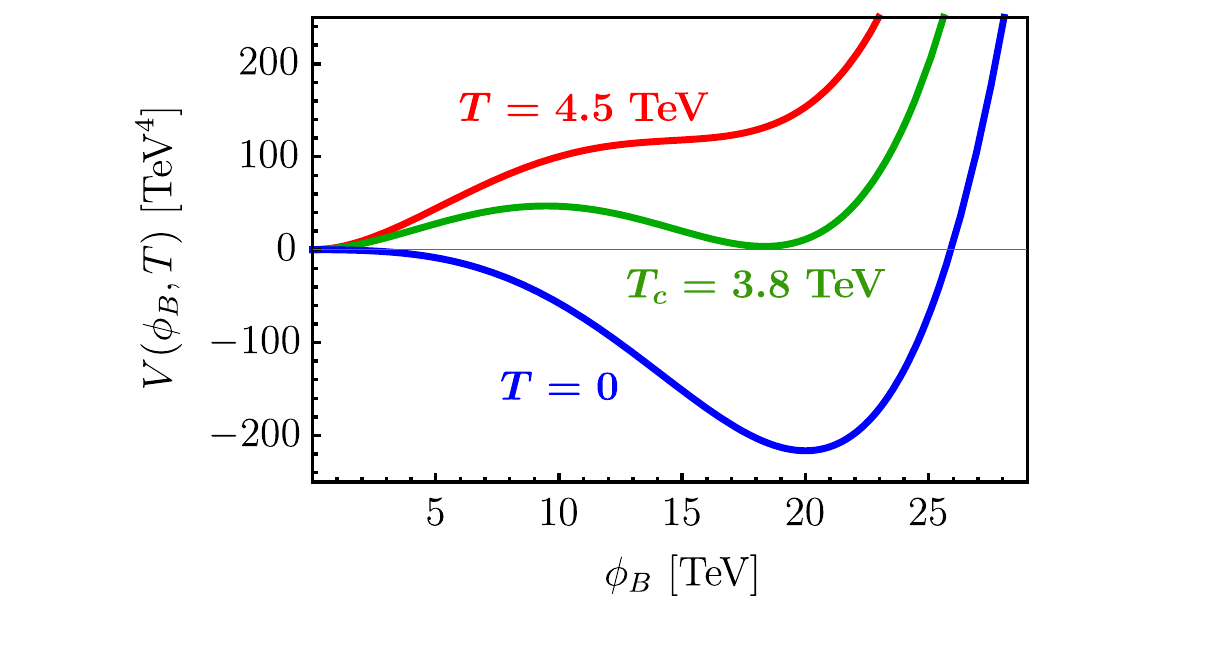} \vspace{-0mm}
\caption{Plot of the  effective potential $V_{\rm eff}(\phi_B, T) - V_{\rm eff}(0, T)$ for the choice of parameters in Eq.\,({\ref{bench}}) and several temperatures.\\}\label{Peff}
\end{figure}

\vspace{1mm}
\noindent
The finite temperature part of the effective potential is 
\bea
&&V_{\rm temp}(\phi_B, T) \nn\\[2pt]
&&= \frac{T^4}{2\pi^2} \sum_i n_i \int_0^\infty dy \,y^2 \log\left(1\mp e^{-\sqrt{m_i^2(\phi_B)/T^2 + y^2}}\right)\nn\\
&& +\, \frac{T}{12\pi} \sum_j n'_j \Big\{m_j^3(\phi_B) - [m_j^2(\phi_B) + \Pi_j(T)]^{\frac32}\Big\} \ .
\eea
In the expression above, the sum over $i$ includes all particles coupling to $\phi_B$, whereas that over $j$ includes only bosons. The plus/minus sign corresponds to fermions/bosons. Ignoring the terms suppressed by small $\lambda_B$, the  thermal masses  are
 \bea
 \Pi_{\phi_B}(T) = \Pi_{\chi_B}(T) =  \tfrac94 g_B^2 T^2 , \ \ \   \Pi_{Z_B}^L(T) = \tfrac{14}{3}g_B^2 T^2  , \ \ \ \ \ \ \ 
 \eea
where the superscript $L$ denotes 
longitudinal components.
\vspace{1mm}

\noindent
The full effective potential is given by
\bea
V_{\rm eff}(\phi_B, T) = V_{\rm tree}(\phi_B) + V_{\rm 1\text{-}loop}(\phi_B) + V_{\rm temp}(\phi_B, T) \, .\nn\\
\eea
It  is shown in Fig.\,\ref{Peff} for the benchmark scenario in Eq.\,(\ref{bench}). A strong first order phase transition occurs, since there is a barrier separating the false vacuum from the true one.

\subsection{Phase transition}

When a patch of the universe tunnels from the false vacuum to the true vacuum, a bubble is formed and starts expanding. The  nucleation rate of such bubbles per unit volume is \cite{LINDE1983421}
\bea\label{gamma}
\Gamma(T) \,\sim \, T^4 \exp\left(-\frac{S(T)}{T}\right) \, ,
\eea
where the Euclidean action $S(T)$ is given by
\bea
S(T) = 4\pi \int dr \, r^2 \left[\,\frac12\, \phi_b'(r)^2 + V_{\rm eff}(\phi_b, T)\,\right]  \ \ \ 
\eea
and $\phi(r)$ is the solution of the expanding bubble equation with appropriate boundary conditions,
\bea
&&\phi_b''(r) + \frac2{r} \,\phi_b'(r) - \frac{d V_{\rm eff}(\phi, T)}{d\phi}\bigg|_{\phi=\phi_b} = 0 \ , \nn\\
&& \phi_b'(0) = 0  \ , \ \ \ \ \ \phi_b(\infty) = \phi_{\rm true}\,. 
\eea
The onset of the phase transition occurs at the nucleation temperature $T_*$ at which $\Gamma(T_*) \approx H^4$.
Using Eq.\,(\ref{gamma}), this condition can be rewritten as
\bea
4\,\log\left(\frac{M_P}{T_*}\right) \approx \frac{S(T_*)}{T_*} \ .
\eea

A phase transition is fully described by four parameters: the bubble wall velocity $v_w$, the nucleation temperature $T_*$, the inverse of its duration $\tilde\beta$,
\bea
\tilde\beta  = T_* \frac{d}{dT}\left(\frac{S(T)}{T}\right)\bigg|_{T=T_*} \ ,
\eea
and the  strength of the transition $\alpha$,
\bea
\alpha = \frac{\rho_{\rm vac}(T_*)}{\rho_{\rm rad}(T_*)} \ .
\eea
In the expression above
\bea
\rho_{\rm vac}(T_*) &=& V_{\rm eff}(\phi_{\rm false},T_*) - V_{\rm eff}(\phi_{\rm true},T_*)\nn\\
&-& T_*\frac{\partial\,[\,V_{\rm eff}(\phi_{\rm false},T) - V_{\rm eff}(\phi_{\rm true},T)\,]}{\partial T}\bigg|_{T=T_*}  \ \ \ \ \ \ \ \ 
\eea
is the energy density of the false vacuum and 
\bea
\rho_{\rm rad}(T_*) = \frac{\pi^2}{30}\,g_*(T_*) \,T_*^4
\eea
is the radiation energy density, with $g_*(T_*)$ being the number of relativistic degrees of freedom at the time of the transition. 

Out of the four parameters $v_w$, $T_*$, $\tilde\beta$, $\alpha$, only the bubble wall velocity  does not depend on the shape of the effective\break potential, and we will set it to $v_w \approx 0.7\,c$ (for a detailed discussion of the possible choices see \cite{Espinosa:2010hh}). For the benchmark scenario in Eq.\,(\ref{bench}) the nucleation temperature is $T_* \approx 600 \ {\rm GeV}$  and  $g_*(T_*) \approx 107$, since all degrees of freedom\break beyond the Standard Model are nonrelativistic at this $T_*$.

\subsection{Gravitational wave signal}

The sound wave contribution to the gravitational wave spectrum  is given by  \cite{Hindmarsh:2013xza,Caprini:2015zlo}
\bea
h^2 \Omega_{s}(\nu) &\,\approx\,& (1.86 \times 10^{-5}) \, \frac{\big(\frac{\nu}{\nu_s}\big)^3}{\left[1+0.75\,\big(\frac{\nu}{\nu_s}\big)^2\right]^{\frac72}}\nn\\
&\times& \, \frac{v_w}{\tilde\beta}\left(\frac{\kappa_s \,\alpha}{\alpha+1}\right)^2 \left(\frac{100}{g_*}\right)^{\frac13} \Upsilon \ ,
\eea
where the parameter $\kappa_s$  (the fraction of the latent heat transformed into the bulk motion of the plasma \cite{Espinosa:2010hh})  and the peak frequency $\nu_s$ are 
\bea
\kappa_s &\approx& \frac{\alpha}{0.73+0.083\sqrt\alpha+\alpha} \ ,\nn\\
\nu_s &\approx& (1.9\times10^{-4} \ {\rm Hz})\left(\frac{g_*}{100}\right)^{\frac16}\frac{\tilde\beta}{v_w} \left(\frac{T_*}{1 \ {\rm TeV}}\right)  \, ,
\eea
and $\Upsilon$ is the suppression factor adopted from \cite{Guo:2020grp},
\bea
\Upsilon = 1- \left[1+\frac{8\pi^{\frac13}}{\sqrt3}\,\frac{v_w}{\tilde{\beta}}\left(\frac{\alpha\,\kappa_s}{\alpha+1}\right)^{\!\!-\frac12}\right]^{-\frac12}  . 
\eea 
This leads to a stronger suppression of the sound wave signal than  previously estimated \cite{Hindmarsh:2017gnf,Ellis:2018mja}, weakening it by nearly two orders of magnitude. 
\vspace{1mm}

\begin{figure}[t!]
\includegraphics[trim={1.0cm 0.5cm 1.2cm 0},clip,width=9.0cm]{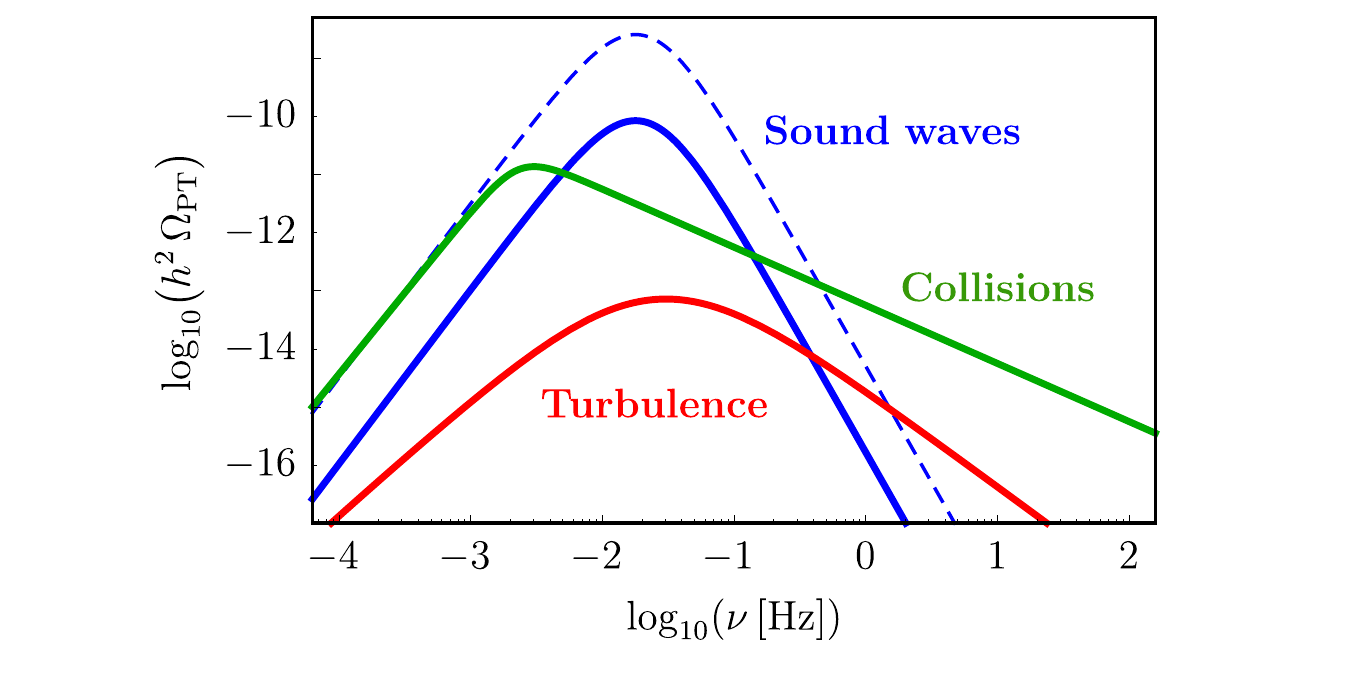} \vspace{-4mm}
\caption{Stochastic gravitational wave  background produced by sounds waves, bubble collisions and turbulence during a first order phase transition triggered by the breaking of gauged ${\rm U}(1)_B$ for the benchmark scenario in Eq.\,(\ref{bench}). The dashed line corresponds to the signal from  sound waves without the  suppression factor $\Upsilon$.\\}\label{fPT}
\end{figure}

\begin{figure*}[t!]
\includegraphics[trim={0cm 0.0cm 0 0},clip,width=15cm]{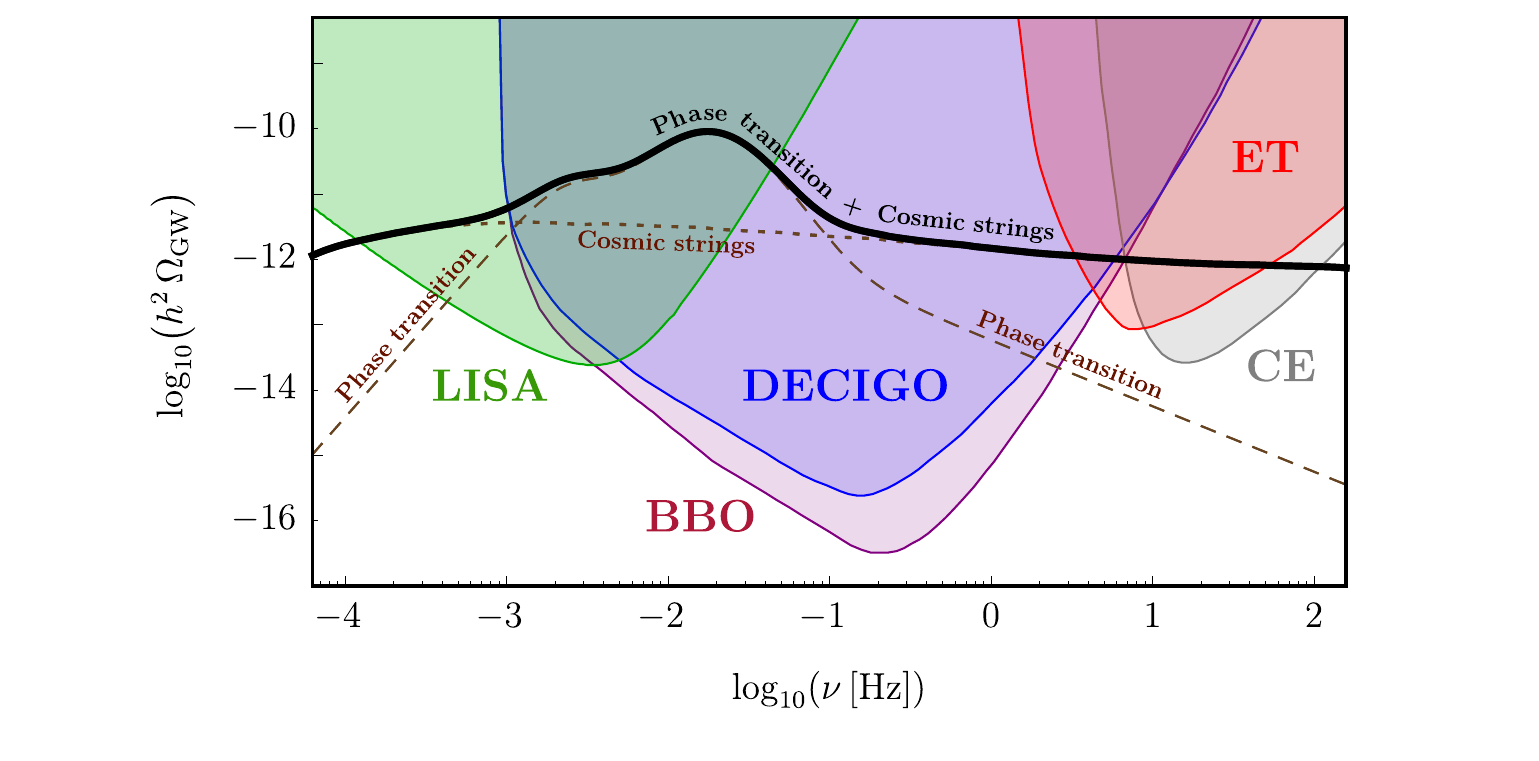} \vspace{-3mm}
\caption{Stochastic gravitational wave signature of the model with gauged baryon and lepton number for $v_B = 20 \ {\rm TeV}$ and $v_L = 10^{11} \ {\rm GeV}$ (black solid line). Sensitivities of future detectors are also shown: 
Big Bang Observer \cite{Yagi:2011wg} (purple), DECIGO \cite{Yagi:2011wg} (blue), LISA (in the C1 configuration)  \cite{Caprini:2015zlo} (green), Cosmic Explorer \cite{Reitze:2019iox} (gray) and Einstein Telescope \cite{Sathyaprakash:2012jk} (red). The  dashed  line denotes the contribution from the phase transition associated with ${\rm U}(1)_B$ breaking, whereas the dotted line corresponds to the cosmic string signal from  ${\rm U}(1)_L$ breaking.\vspace{7mm}}\label{fig4}
\end{figure*}

\noindent
The contribution to the gravitational wave spectrum arising from bubble collisions is \cite{Kosowsky:1991ua,Huber:2008hg,Caprini:2015zlo}
\bea
h^2 \Omega_{c}(\nu) &\,\approx\,& (1.66 \times 10^{-5}) \, \frac{\big(\frac{\nu}{\nu_c}\big)^{2.8}}{1+2.8\,\big(\frac{\nu}{\nu_c}\big)^{3.8}}\nn\\
&\times&\! \left(\frac{v_w^3}{1+2.4\,v_w^2}\right)\frac{1}{\tilde\beta^2}\left(\frac{\kappa_c \,\alpha}{\alpha+1}\right)^2 \left(\frac{100}{g_*}\right)^{\frac13} \! , \ \ \ \ \ \ \ 
\eea
where $\kappa_c$ (the fraction of the latent heat deposited into the bubble front  \cite{Kamionkowski:1993fg}) and the peak frequency $\nu_c$ are
\bea
\kappa_c &\approx& \frac{0.715\,\alpha+\frac{4}{27}\sqrt{\frac{3\alpha}{2}}}{1+0.715\alpha} \ ,\\
\nu_c &\approx& (10^{-4} \ {\rm Hz})\left(\frac{g_*}{100}\right)^{\!\frac16}\!\left(\frac{\tilde\beta}{1.8-0.1v_w+v_w^2}\right) \!\left(\frac{T_*}{1 \ {\rm TeV}}\right) \! .  \nn
\eea
Finally, the contribution from turbulence is  \cite{Kamionkowski:1993fg}
\bea
h^2 \Omega_{t}(\nu) &\,\approx\,& (3.35 \times 10^{-4}) \, \frac{\big(\frac{\nu}{\nu_t}\big)^{3}}{\big(1+\frac{8\pi \nu}{h_*}\big)\big(1+\frac{\nu}{\nu_t}\big)^{\frac{11}{3}}}\nn\\
&\times&  \frac{v_w}{\tilde\beta}\left(\frac{\epsilon \,\kappa_s \,\alpha}{\alpha+1}\right)^{\frac32} \left(\frac{100}{g_*}\right)^{\frac13} \! , \ \ \ \ \ 
\eea
where we adopt   $\epsilon = 0.05$ \cite{Caprini:2015zlo} and 
\bea
h_* &=& (1.7\times 10^{-4} \ {\rm Hz})\left(\frac{g_*}{100}\right)^{\frac16} \left(\frac{T_*}{1 \ {\rm TeV}}\right) \ ,\nn\\
\nu_t &=& 1.64 \, \frac{\tilde\beta}{v_w} \, h_*  \ . 
\eea

Figure \ref{fPT} shows the  individual contributions to the stochastic gravitational wave background  from  ${\rm U}(1)_B$ breaking that originate from sound waves, bubble collisions and magnetohydrodynamic turbulence, for the benchmark scenario in Eq.\,(\ref{bench}) and in the frequency range relevant for upcoming gravitational wave experiments. 
Note that without  the suppression factor the signal from sound waves would be dominant, but  including this factor results in the sound wave and bubble collision contributions being of  similar magnitude.

\section{Gravitational wave signature}

Figure  \ref{fig4} shows the combined gravitational wave signature of the cosmic string network produced at the high scale and a first order phase transition that occurred  at the low scale. The plot was made for the model with gauged baryon and lepton number with  $v_L = 10^{11} \ {\rm GeV}$ and the  parameter values as in Eq.\,(\ref{bench}).
The expected signal is flat throughout a wide range of frequencies and contains a characteristic bump-like feature, which distinguishes it from  pure seesaw signatures. 

The position of the bump depends linearly on the nucleation temperature; for a higher ${\rm U}(1)_B$ breaking scale the phase transition appears at higher frequencies. 
The parameter $\tilde\beta$ affects both the position of the peak and its height, whereas the parameter $\alpha$  governs only its height; a larger $\tilde\beta$ corresponds to higher frequencies and a weaker signal, whereas a larger $\alpha$  implies a stronger signal.  Upon implementing the theoretically predicted suppression of the sound wave signal, a double-bump feature emerges from the competition between the sound wave and bubble collision contributions.

The breadth of the signature across  many frequencies places it within the reach of nearly all upcoming  gravitational wave detectors. The high-frequency flat part of the signal  can be searched for in experiments like the Cosmic Explorer and the Einstein Telescope, whereas  the low-frequency part including the bump feature is within the reach  of LISA,  the Big Bang Explorer and DECIGO. The signal is clearly distinguishable  from a pure cosmic string signature (shown as the brown dotted line) and from a pure phase transition signature (denoted by the brown dashed line).

\section{Summary}

We have recently entered an extremely exciting time when progress in particle physics may actually come from classical gravity measurements.  Gravitational wave detectors offer a  very  promising  probe of the early universe and may  provide information on the structure of the theory at high scales, well  above the LHC reach and inaccessible directly in any other existing experiment. 
There are generally two kinds of particle\break physics signatures which can be searched for via gravitational wave measurements, and that fall within the sensitivity of near-future gravitational wave experiments. 
\vspace{1mm}

Signals of the first type arise from cosmic phase transitions, and are produced abruptly by sound waves, bubble collisions and magnetohydrodynamic turbulence. They exhibit a bump-like shape, with the  peak frequency determined by the symmetry breaking scale. 
The second class of signals comes from the dynamics of the cosmic string network, produced during a phase transition, but sourcing gravitational radiation throughout  a long period after its formation.  Those signals are flat and stretch out across a wide range of frequencies.
\vspace{1mm}

In this paper, we considered the possibility of the two types of signals co-existing and giving rise to a new type of signature --  a flat spectrum with a bump feature --  which is within the reach of upcoming gravitational wave experiments. We pointed out that such signatures occur generically in models with two or more gauge symmetries that are broken at vastly separated scales. We  analyzed this scenario in the context of a model with gauged baryon and lepton number, where the high-scale breaking of lepton number is motivated by the seesaw mechanism for the neutrinos, whereas the  breaking of baryon number is confined to   a much lower scale by the\break observed dark matter relic abundance. The flat part of the\break resulting spectrum is within the reach of the Cosmic Explorer and Einstein Telescope, whereas the bump feature falls  within the sensitivity of LISA, Big Bang Observer and DECIGO.
\vspace{1mm}

Such a cosmic search for a combined signature  of baryon and lepton number violation is complementary to collider\break efforts.  An observation of the gravitational wave  signal proposed here would be a strong motivation for building the $100 \ {\rm TeV}$ collider, which could independently  search for the leptophobic gauge boson associated with baryon number breaking. 
Such a discovery would also imply the necessity of revisiting the  ideas about grand unification, once again\break showing  that nature is full of surprises.

\subsection*{Acknowledgments}

The authors are grateful to Yue Zhao for illuminating\break discussions and valuable comments.  The work of B.F. was supported in part by the U.S. Department of Energy under Award
No.~${\rm DE}$-${\rm SC0009959}$. The work of B.S. was supported in part by the NSF grant ${\rm PHY}$-${\rm 1720282}$.

\bibliography{PeV}

\begin{thebibliography}{84}%
\makeatletter
\providecommand \@ifxundefined [1]{%
 \@ifx{#1\undefined}
}%
\providecommand \@ifnum [1]{%
 \ifnum #1\expandafter \@firstoftwo
 \else \expandafter \@secondoftwo
 \fi
}%
\providecommand \@ifx [1]{%
 \ifx #1\expandafter \@firstoftwo
 \else \expandafter \@secondoftwo
 \fi
}%
\providecommand \natexlab [1]{#1}%
\providecommand \enquote  [1]{#1}%
\providecommand \bibnamefont  [1]{#1}%
\providecommand \bibfnamefont [1]{#1}%
\providecommand \citenamefont [1]{#1}%
\providecommand \href@noop [0]{\@secondoftwo}%
\providecommand \href [0]{\begingroup \@sanitize@url \@href}%
\providecommand \@href[1]{\@@startlink{#1}\@@href}%
\providecommand \@@href[1]{\endgroup#1\@@endlink}%
\providecommand \@sanitize@url [0]{\catcode `\\12\catcode `\$12\catcode
  `\&12\catcode `\#12\catcode `\^12\catcode `\_12\catcode `\%12\relax}%
\providecommand \@@startlink[1]{}%
\providecommand \@@endlink[0]{}%
\providecommand \url  [0]{\begingroup\@sanitize@url \@url }%
\providecommand \@url [1]{\endgroup\@href {#1}{\urlprefix }}%
\providecommand \urlprefix  [0]{URL }%
\providecommand \Eprint [0]{\href }%
\providecommand \doibase [0]{http://dx.doi.org/}%
\providecommand \selectlanguage [0]{\@gobble}%
\providecommand \bibinfo  [0]{\@secondoftwo}%
\providecommand \bibfield  [0]{\@secondoftwo}%
\providecommand \translation [1]{[#1]}%
\providecommand \BibitemOpen [0]{}%
\providecommand \bibitemStop [0]{}%
\providecommand \bibitemNoStop [0]{.\EOS\space}%
\providecommand \EOS [0]{\spacefactor3000\relax}%
\providecommand \BibitemShut  [1]{\csname bibitem#1\endcsname}%
\let\auto@bib@innerbib\@empty
%</preamble>
\bibitem [{\citenamefont {Aasi}\ \emph {et~al.}(2015)\citenamefont {Aasi} \emph
  {et~al.}}]{TheLIGOScientific:2014jea}%
  \BibitemOpen
  \bibfield  {author} {\bibinfo {author} {\bibfnamefont {J.}~\bibnamefont
  {Aasi}} \emph {et~al.} (\bibinfo {collaboration} {LIGO Scientific}),\
  }\bibfield  {title} {\enquote {\bibinfo {title} {\emph{Advanced LIGO}},}\
  }\href {\doibase 10.1088/0264-9381/32/7/074001} {\bibfield  {journal}
  {\bibinfo  {journal} {Class. Quant. Grav.}\ }\textbf {\bibinfo {volume}
  {32}},\ \bibinfo {pages} {074001} (\bibinfo {year} {2015})},\ \Eprint
  {http://arxiv.org/abs/1411.4547} {arXiv:1411.4547 [gr-qc]} \BibitemShut
  {NoStop}%
\bibitem [{\citenamefont {Acernese}\ \emph {et~al.}(2015)\citenamefont
  {Acernese} \emph {et~al.}}]{TheVirgo:2014hva}%
  \BibitemOpen
  \bibfield  {author} {\bibinfo {author} {\bibfnamefont {F.}~\bibnamefont
  {Acernese}} \emph {et~al.} (\bibinfo {collaboration} {VIRGO}),\ }\bibfield
  {title} {\enquote {\bibinfo {title} {\emph{Advanced Virgo: A
  Second-Generation Interferometric Gravitational Wave Detector}},}\ }\href
  {\doibase 10.1088/0264-9381/32/2/024001} {\bibfield  {journal} {\bibinfo
  {journal} {Class. Quant. Grav.}\ }\textbf {\bibinfo {volume} {32}},\ \bibinfo
  {pages} {024001} (\bibinfo {year} {2015})},\ \Eprint
  {http://arxiv.org/abs/1408.3978} {arXiv:1408.3978 \ \ \ \ \ \ \ [gr-qc]} \BibitemShut
  {NoStop}%
\bibitem [{\citenamefont {Amaro-Seoane}\ \emph {et~al.}(2017)\citenamefont
  {Amaro-Seoane} \emph {et~al.}}]{Audley:2017drz}%
  \BibitemOpen
  \bibfield  {author} {\bibinfo {author} {\bibfnamefont {P.}~\bibnamefont
  {Amaro-Seoane}} \emph {et~al.} (\bibinfo {collaboration} {LISA}),\ }\bibfield
   {title} {\enquote {\bibinfo {title} {\emph{Laser Interferometer Space
  Antenna}},}\ }\href@noop {} {(\bibinfo {year} {2017})},\ \Eprint
  {http://arxiv.org/abs/1702.00786} {arXiv:1702.00786 [astro-ph.IM]}
  \BibitemShut {NoStop}%
\bibitem [{\citenamefont {Kawamura}\ \emph {et~al.}(2011)\citenamefont
  {Kawamura} \emph {et~al.}}]{Kawamura:2011zz}%
  \BibitemOpen
  \bibfield  {author} {\bibinfo {author} {\bibfnamefont {S.}~\bibnamefont
  {Kawamura}} \emph {et~al.},\ }\bibfield  {title} {\enquote {\bibinfo {title}
  {\emph{The Japanese space gravitational wave antenna: DECIGO}},}\ }\href
  {\doibase 10.1088/0264-9381/28/9/094011} {\bibfield  {journal} {\bibinfo
  {journal} {Class. Quant. Grav.}\ }\textbf {\bibinfo {volume} {28}},\ \bibinfo
  {pages} {094011} (\bibinfo {year} {2011})}\BibitemShut {NoStop}%
\bibitem [{\citenamefont {Reitze}\ \emph {et~al.}(2019)\citenamefont {Reitze}
  \emph {et~al.}}]{Reitze:2019iox}%
  \BibitemOpen
  \bibfield  {author} {\bibinfo {author} {\bibfnamefont {D.}~\bibnamefont
  {Reitze}} \emph {et~al.},\ }\bibfield  {title} {\enquote {\bibinfo {title}
  {\emph{Cosmic Explorer: The U.S. Contribution to Gravitational-Wave Astronomy
  beyond LIGO}},}\ }\href@noop {} {\bibfield  {journal} {\bibinfo  {journal}
  {Bull. Am. Astron. Soc.}\ }\textbf {\bibinfo {volume} {51}},\ \bibinfo
  {pages} {035} (\bibinfo {year} {2019})},\ \Eprint
  {http://arxiv.org/abs/1907.04833} {arXiv:1907.04833 [astro-ph.IM]}
  \BibitemShut {NoStop}%
\bibitem [{\citenamefont {Crowder} and\ \citenamefont
  {Cornish}(2005)}]{Crowder:2005nr}%
  \BibitemOpen
  \bibfield  {author} {\bibinfo {author} {\bibfnamefont {J.}~\bibnamefont
  {Crowder}} and\ \bibinfo {author} {\bibfnamefont {N.~J.}\ \bibnamefont
  {Cornish}},\ }\bibfield  {title} {\enquote {\bibinfo {title} {\emph{Beyond
  LISA: Exploring Future Gravitational Wave Missions}},}\ }\href {\doibase
  10.1103/PhysRevD.72.083005} {\bibfield  {journal} {\bibinfo  {journal} {Phys.
  Rev. D}\ }\textbf {\bibinfo {volume} {72}},\ \bibinfo {pages} {083005}
  (\bibinfo {year} {2005})},\ \Eprint {http://arxiv.org/abs/gr-qc/0506015}
  {arXiv:gr-qc/0506015} \BibitemShut {NoStop}%
\bibitem [{\citenamefont {Apreda}\ \emph {et~al.}(2002)\citenamefont {Apreda},
  \citenamefont {Maggiore}, \citenamefont {Nicolis}, and\ \citenamefont
  {Riotto}}]{Apreda:2001us}%
  \BibitemOpen
  \bibfield  {author} {\bibinfo {author} {\bibfnamefont {R.}~\bibnamefont
  {Apreda}}, \bibinfo {author} {\bibfnamefont {M.}~\bibnamefont {Maggiore}},
  \bibinfo {author} {\bibfnamefont {A.}~\bibnamefont {Nicolis}},  and\
  \bibinfo {author} {\bibfnamefont {A.}~\bibnamefont {Riotto}},\ }\bibfield
  {title} {\enquote {\bibinfo {title} {\emph{Gravitational Waves from
  Electroweak Phase Transitions}},}\ }\href {\doibase
  10.1016/S0550-3213(02)00264-X} {\bibfield  {journal} {\bibinfo  {journal}
  {Nucl. Phys. B}\ }\textbf {\bibinfo {volume} {631}},\ \bibinfo {pages}
  {342--368} (\bibinfo {year} {2002})},\ \Eprint
  {http://arxiv.org/abs/gr-qc/0107033} {arXiv:gr-qc/0107033} \BibitemShut
  {NoStop}%
\bibitem [{\citenamefont {Grojean} and\ \citenamefont
  {Servant}(2007)}]{Grojean:2006bp}%
  \BibitemOpen
  \bibfield  {author} {\bibinfo {author} {\bibfnamefont {C.}~\bibnamefont
  {Grojean}} and\ \bibinfo {author} {\bibfnamefont {G.}~\bibnamefont
  {Servant}},\ }\bibfield  {title} {\enquote {\bibinfo {title}
  {\emph{Gravitational Waves from Phase Transitions at the Electroweak Scale
  and Beyond}},}\ }\href {\doibase 10.1103/PhysRevD.75.043507} {\bibfield
  {journal} {\bibinfo  {journal} {Phys. Rev. D}\ }\textbf {\bibinfo {volume}
  {75}},\ \bibinfo {pages} {043507} (\bibinfo {year} {2007})},\ \Eprint
  {http://arxiv.org/abs/hep-ph/0607107} {arXiv:hep-ph/0607107} \BibitemShut
  {NoStop}%
\bibitem [{\citenamefont {Leitao}\ \emph {et~al.}(2012)\citenamefont {Leitao},
  \citenamefont {Megevand}, and\ \citenamefont {Sanchez}}]{Leitao:2012tx}%
  \BibitemOpen
  \bibfield  {author} {\bibinfo {author} {\bibfnamefont {L.}~\bibnamefont
  {Leitao}}, \bibinfo {author} {\bibfnamefont {A.}~\bibnamefont {Megevand}}, 
  and\ \bibinfo {author} {\bibfnamefont {A.~D.}\ \bibnamefont {Sanchez}},\
  }\bibfield  {title} {\enquote {\bibinfo {title} {\emph{Gravitational Waves
  from the Electroweak Phase Transition}},}\ }\href {\doibase
  10.1088/1475-7516/2012/10/024} {\bibfield  {journal} {\bibinfo  {journal}
  {JCAP}\ }\textbf {\bibinfo {volume} {10}},\ \bibinfo {pages} {024} (\bibinfo
  {year} {2012})},\ \Eprint {http://arxiv.org/abs/1205.3070} {arXiv:1205.3070
  [astro-ph.CO]} \BibitemShut {NoStop}%
\bibitem [{\citenamefont {Schwaller}(2015)}]{Schwaller:2015tja}%
  \BibitemOpen
  \bibfield  {author} {\bibinfo {author} {\bibfnamefont {P.}~\bibnamefont
  {Schwaller}},\ }\bibfield  {title} {\enquote {\bibinfo {title}
  {\emph{Gravitational Waves from a Dark Phase Transition}},}\ }\href {\doibase
  10.1103/PhysRevLett.115.181101} {\bibfield  {journal} {\bibinfo  {journal}
  {Phys. Rev. Lett.}\ }\textbf {\bibinfo {volume} {115}},\ \bibinfo {pages}
  {181101} (\bibinfo {year} {2015})},\ \Eprint
  {http://arxiv.org/abs/1504.07263} {arXiv:1504.07263 [hep-ph]} \BibitemShut
  {NoStop}%
\bibitem [{\citenamefont {Huang} and\ \citenamefont
  {Zhang}(2019)}]{Huang:2017laj}%
  \BibitemOpen
  \bibfield  {author} {\bibinfo {author} {\bibfnamefont {F.~P.}\ \bibnamefont
  {Huang}} and\ \bibinfo {author} {\bibfnamefont {X.}~\bibnamefont {Zhang}},\
  }\bibfield  {title} {\enquote {\bibinfo {title} {\emph{Probing the Gauge
  Symmetry Breaking of the Early Universe in 3-3-1 Models and Beyond by
  Gravitational Waves}},}\ }\href {\doibase 10.1016/j.physletb.2018.11.024}
  {\bibfield  {journal} {\bibinfo  {journal} {Phys. Lett. B}\ }\textbf
  {\bibinfo {volume} {788}},\ \bibinfo {pages} {288--294} (\bibinfo {year}
  {2019})},\ \Eprint {http://arxiv.org/abs/1701.04338} {arXiv:1701.04338
  [hep-ph]} \BibitemShut {NoStop}%
\bibitem [{\citenamefont {Huang} and\ \citenamefont
  {Yu}(2018)}]{Huang:2017rzf}%
  \BibitemOpen
  \bibfield  {author} {\bibinfo {author} {\bibfnamefont {F.~P.}\ \bibnamefont
  {Huang}} and\ \bibinfo {author} {\bibfnamefont {Jiang-Hao}\ \bibnamefont
  {Yu}},\ }\bibfield  {title} {\enquote {\bibinfo {title} {\emph{Exploring
  Inert Dark Matter Blind Spots with Gravitational Wave Signatures}},}\ }\href
  {\doibase 10.1103/PhysRevD.98.095022} {\bibfield  {journal} {\bibinfo
  {journal} {Phys. Rev. D}\ }\textbf {\bibinfo {volume} {98}},\ \bibinfo
  {pages} {095022} (\bibinfo {year} {2018})},\ \Eprint
  {http://arxiv.org/abs/1704.04201} {arXiv:1704.04201 [hep-ph]} \BibitemShut
  {NoStop}%
\bibitem [{\citenamefont {Demidov}\ \emph {et~al.}(2018)\citenamefont
  {Demidov}, \citenamefont {Gorbunov}, and\ \citenamefont
  {Kirpichnikov}}]{Demidov:2017lzf}%
  \BibitemOpen
  \bibfield  {author} {\bibinfo {author} {\bibfnamefont {S.~V.}\ \bibnamefont
  {Demidov}}, \bibinfo {author} {\bibfnamefont {D.~S.}\ \bibnamefont
  {Gorbunov}},  and\ \bibinfo {author} {\bibfnamefont {D.~V.}\ \bibnamefont
  {Kirpichnikov}},\ }\bibfield  {title} {\enquote {\bibinfo {title}
  {\emph{Gravitational Waves from Phase Transition in Split NMSSM}},}\ }\href
  {\doibase 10.1016/j.physletb.2018.02.007} {\bibfield  {journal} {\bibinfo
  {journal} {Phys. Lett. B}\ }\textbf {\bibinfo {volume} {779}},\ \bibinfo
  {pages} {191--194} (\bibinfo {year} {2018})},\ \Eprint
  {http://arxiv.org/abs/1712.00087} {arXiv:1712.00087 [hep-ph]} \BibitemShut
  {NoStop}%
\bibitem [{\citenamefont {Hashino}\ \emph {et~al.}(2018)\citenamefont
  {Hashino}, \citenamefont {Kakizaki}, \citenamefont {Kanemura}, \citenamefont
  {Ko}, and\ \citenamefont {Matsui}}]{Hashino:2018zsi}%
  \BibitemOpen
  \bibfield  {author} {\bibinfo {author} {\bibfnamefont {K.}~\bibnamefont
  {Hashino}}, \bibinfo {author} {\bibfnamefont {M.}~\bibnamefont {Kakizaki}},
  \bibinfo {author} {\bibfnamefont {S.}~\bibnamefont {Kanemura}}, \bibinfo
  {author} {\bibfnamefont {P.}~\bibnamefont {Ko}},  and\ \bibinfo {author}
  {\bibfnamefont {T.}~\bibnamefont {Matsui}},\ }\bibfield  {title} {\enquote
  {\bibinfo {title} {\emph{Gravitational Waves from First Order Electroweak
  Phase Transition in Models with the U(1)$_{X}$ Gauge Symmetry}},}\ }\href
  {\doibase 10.1007/JHEP06(2018)088} {\bibfield  {journal} {\bibinfo  {journal}
  {JHEP}\ }\textbf {\bibinfo {volume} {06}},\ \bibinfo {pages} {088} (\bibinfo
  {year} {2018})},\ \Eprint {http://arxiv.org/abs/1802.02947} {arXiv:1802.02947
  [hep-ph]} \BibitemShut {NoStop}%
\bibitem [{\citenamefont {Okada} and\ \citenamefont
  {Seto}(2018)}]{Okada:2018xdh}%
  \BibitemOpen
  \bibfield  {author} {\bibinfo {author} {\bibfnamefont {N.}~\bibnamefont
  {Okada}} and\ \bibinfo {author} {\bibfnamefont {O.}~\bibnamefont {Seto}},\
  }\bibfield  {title} {\enquote {\bibinfo {title} {\emph{Probing the Seesaw
  Scale with Gravitational Waves}},}\ }\href {\doibase
  10.1103/PhysRevD.98.063532} {\bibfield  {journal} {\bibinfo  {journal} {Phys.
  Rev. D}\ }\textbf {\bibinfo {volume} {98}},\ \bibinfo {pages} {063532}
  (\bibinfo {year} {2018})},\ \Eprint {http://arxiv.org/abs/1807.00336}
  {arXiv:1807.00336 [hep-ph]} \BibitemShut {NoStop}%
\bibitem [{\citenamefont {Ahriche}\ \emph {et~al.}(2019)\citenamefont
  {Ahriche}, \citenamefont {Hashino}, \citenamefont {Kanemura}, and\
  \citenamefont {Nasri}}]{Ahriche:2018rao}%
  \BibitemOpen
  \bibfield  {author} {\bibinfo {author} {\bibfnamefont {A.}~\bibnamefont
  {Ahriche}}, \bibinfo {author} {\bibfnamefont {K.}~\bibnamefont {Hashino}},
  \bibinfo {author} {\bibfnamefont {S.}~\bibnamefont {Kanemura}},  and\
  \bibinfo {author} {\bibfnamefont {S.}~\bibnamefont {Nasri}},\ }\bibfield
  {title} {\enquote {\bibinfo {title} {\emph{Gravitational Waves from Phase
  Transitions in Models with Charged Singlets}},}\ }\href {\doibase
  10.1016/j.physletb.2018.12.013} {\bibfield  {journal} {\bibinfo  {journal}
  {Phys. Lett. B}\ }\textbf {\bibinfo {volume} {789}},\ \bibinfo {pages}
  {119--126} (\bibinfo {year} {2019})},\ \Eprint
  {http://arxiv.org/abs/1809.09883} {arXiv:1809.09883 [hep-ph]} \BibitemShut
  {NoStop}%
\bibitem [{\citenamefont {Brdar}\ \emph
  {et~al.}(2019{\natexlab{a}})\citenamefont {Brdar}, \citenamefont
  {Helmboldt}, and\ \citenamefont {Kubo}}]{Brdar:2018num}%
  \BibitemOpen
  \bibfield  {author} {\bibinfo {author} {\bibfnamefont {V.}~\bibnamefont
  {Brdar}}, \bibinfo {author} {\bibfnamefont {A.~J.}\ \bibnamefont
  {Helmboldt}},  and\ \bibinfo {author} {\bibfnamefont {J.}~\bibnamefont
  {Kubo}},\ }\bibfield  {title} {\enquote {\bibinfo {title}
  {\emph{Gravitational Waves from First-Order Phase Transitions: LIGO as a
  Window to Unexplored Seesaw Scales}},}\ }\href {\doibase
  10.1088/1475-7516/2019/02/021} {\bibfield  {journal} {\bibinfo  {journal}
  {JCAP}\ }\textbf {\bibinfo {volume} {02}},\ \bibinfo {pages} {021} (\bibinfo
  {year} {2019}{\natexlab{a}})},\ \Eprint {http://arxiv.org/abs/1810.12306}
  {arXiv:1810.12306 [hep-ph]} \BibitemShut {NoStop}%
\bibitem [{\citenamefont {Croon}\ \emph {et~al.}(2019)\citenamefont {Croon},
  \citenamefont {Gonzalo}, and\ \citenamefont {White}}]{Croon:2018kqn}%
  \BibitemOpen
  \bibfield  {author} {\bibinfo {author} {\bibfnamefont {D.}~\bibnamefont
  {Croon}}, \bibinfo {author} {\bibfnamefont {T.~E.}\ \bibnamefont {Gonzalo}},
   and\ \bibinfo {author} {\bibfnamefont {G.}~\bibnamefont {White}},\
  }\bibfield  {title} {\enquote {\bibinfo {title} {\emph{Gravitational Waves
  from a Pati-Salam Phase Transition}},}\ }\href {\doibase
  10.1007/JHEP02(2019)083} {\bibfield  {journal} {\bibinfo  {journal} {JHEP}\
  }\textbf {\bibinfo {volume} {02}},\ \bibinfo {pages} {083} (\bibinfo {year}
  {2019})},\ \Eprint {http://arxiv.org/abs/1812.02747} {arXiv:1812.02747
  [hep-ph]} \BibitemShut {NoStop}%
\bibitem [{\citenamefont {Angelescu} and\ \citenamefont
  {Huang}(2019)}]{Angelescu:2018dkk}%
  \BibitemOpen
  \bibfield  {author} {\bibinfo {author} {\bibfnamefont {A.}~\bibnamefont
  {Angelescu}} and\ \bibinfo {author} {\bibfnamefont {P.}~\bibnamefont
  {Huang}},\ }\bibfield  {title} {\enquote {\bibinfo {title} {\emph{Multistep
  Strongly First Order Phase Transitions from New Fermions at the TeV
  Scale}},}\ }\href {\doibase 10.1103/PhysRevD.99.055023} {\bibfield  {journal}
  {\bibinfo  {journal} {Phys. Rev. D}\ }\textbf {\bibinfo {volume} {99}},\
  \bibinfo {pages} {055023} (\bibinfo {year} {2019})},\ \Eprint
  {http://arxiv.org/abs/1812.08293} {arXiv:1812.08293 [hep-ph]} \BibitemShut
  {NoStop}%
\bibitem [{\citenamefont {Hasegawa}\ \emph {et~al.}(2019)\citenamefont
  {Hasegawa}, \citenamefont {Okada}, and\ \citenamefont
  {Seto}}]{Hasegawa:2019amx}%
  \BibitemOpen
  \bibfield  {author} {\bibinfo {author} {\bibfnamefont {T.}~\bibnamefont
  {Hasegawa}}, \bibinfo {author} {\bibfnamefont {N.}~\bibnamefont {Okada}}, \
  and\ \bibinfo {author} {\bibfnamefont {O.}~\bibnamefont {Seto}},\ }\bibfield
  {title} {\enquote {\bibinfo {title} {\emph{Gravitational Waves from the
  Minimal Gauged $U(1)_{B-L}$ Model}},}\ }\href {\doibase
  10.1103/PhysRevD.99.095039} {\bibfield  {journal} {\bibinfo  {journal} {Phys.
  Rev. D}\ }\textbf {\bibinfo {volume} {99}},\ \bibinfo {pages} {095039}
  (\bibinfo {year} {2019})},\ \Eprint {http://arxiv.org/abs/1904.03020}
  {arXiv:1904.03020 [hep-ph]} \BibitemShut {NoStop}%
\bibitem [{\citenamefont {Dev}\ \emph {et~al.}(2019)\citenamefont {Dev},
  \citenamefont {Ferrer}, \citenamefont {Zhang}, and\ \citenamefont
  {Zhang}}]{Dev:2019njv}%
  \BibitemOpen
  \bibfield  {author} {\bibinfo {author} {\bibfnamefont {P.~S.~B.}\
  \bibnamefont {Dev}}, \bibinfo {author} {\bibfnamefont {F.}~\bibnamefont
  {Ferrer}}, \bibinfo {author} {\bibfnamefont {Y.}~\bibnamefont {Zhang}}, \
  and\ \bibinfo {author} {\bibfnamefont {Y.}~\bibnamefont {Zhang}},\ }\bibfield
   {title} {\enquote {\bibinfo {title} {\emph{Gravitational Waves from
  First-Order Phase Transition in a Simple Axion-Like Particle Model}},}\
  }\href {\doibase 10.1088/1475-7516/2019/11/006} {\bibfield  {journal}
  {\bibinfo  {journal} {JCAP}\ }\textbf {\bibinfo {volume} {11}},\ \bibinfo
  {pages} {006} (\bibinfo {year} {2019})},\ \Eprint
  {http://arxiv.org/abs/1905.00891} {arXiv:1905.00891 [hep-ph]} \BibitemShut
  {NoStop}%
\bibitem [{\citenamefont {Brdar}\ \emph
  {et~al.}(2019{\natexlab{b}})\citenamefont {Brdar}, \citenamefont {Graf},
  \citenamefont {Helmboldt}, and\ \citenamefont {Xu}}]{Brdar:2019fur}%
  \BibitemOpen
  \bibfield  {author} {\bibinfo {author} {\bibfnamefont {V.}~\bibnamefont
  {Brdar}}, \bibinfo {author} {\bibfnamefont {L.}~\bibnamefont {Graf}},
  \bibinfo {author} {\bibfnamefont {A.~J.}\ \bibnamefont {Helmboldt}},  and\
  \bibinfo {author} {\bibfnamefont {X.-J.}\ \bibnamefont {Xu}},\ }\bibfield
  {title} {\enquote {\bibinfo {title} {\emph{Gravitational Waves as a Probe of
  Left-Right Symmetry Breaking}},}\ }\href {\doibase
  10.1088/1475-7516/2019/12/027} {\bibfield  {journal} {\bibinfo  {journal}
  {JCAP}\ }\textbf {\bibinfo {volume} {12}},\ \bibinfo {pages} {027} (\bibinfo
  {year} {2019}{\natexlab{b}})},\ \Eprint {http://arxiv.org/abs/1909.02018}
  {arXiv:1909.02018 [hep-ph]} \BibitemShut {NoStop}%
\bibitem [{\citenamefont {Wang}\ \emph {et~al.}(2020)\citenamefont {Wang},
  \citenamefont {Huang}, and\ \citenamefont {Zhang}}]{Wang:2019pet}%
  \BibitemOpen
  \bibfield  {author} {\bibinfo {author} {\bibfnamefont {X.}~\bibnamefont
  {Wang}}, \bibinfo {author} {\bibfnamefont {F.~P.}\ \bibnamefont {Huang}}, \
  and\ \bibinfo {author} {\bibfnamefont {X.}~\bibnamefont {Zhang}},\ }\bibfield
   {title} {\enquote {\bibinfo {title} {\emph{Gravitational Wave and Collider
  Signals in Complex Two-Higgs Doublet Model with Dynamical CP-Violation at
  Finite Temperature}},}\ }\href {\doibase 10.1103/PhysRevD.101.015015}
  {\bibfield  {journal} {\bibinfo  {journal} {Phys. Rev. D}\ }\textbf {\bibinfo
  {volume} {101}},\ \bibinfo {pages} {015015} (\bibinfo {year} {2020})},\
  \Eprint {http://arxiv.org/abs/1909.02978} {arXiv:1909.02978 [hep-ph]}
  \BibitemShut {NoStop}%
\bibitem [{\citenamefont {Greljo}\ \emph {et~al.}(2020)\citenamefont {Greljo},
  \citenamefont {Opferkuch}, and\ \citenamefont {Stefanek}}]{Greljo:2019xan}%
  \BibitemOpen
  \bibfield  {author} {\bibinfo {author} {\bibfnamefont {A.}~\bibnamefont
  {Greljo}}, \bibinfo {author} {\bibfnamefont {T.}~\bibnamefont {Opferkuch}}, \
  and\ \bibinfo {author} {\bibfnamefont {B.~A.}\ \bibnamefont {Stefanek}},\
  }\bibfield  {title} {\enquote {\bibinfo {title} {\emph{Gravitational Imprints
  of Flavor Hierarchies}},}\ }\href {\doibase 10.1103/PhysRevLett.124.171802}
  {\bibfield  {journal} {\bibinfo  {journal} {Phys. Rev. Lett.}\ }\textbf
  {\bibinfo {volume} {124}},\ \bibinfo {pages} {171802} (\bibinfo {year}
  {2020})},\ \Eprint {http://arxiv.org/abs/1910.02014} {arXiv:1910.02014
  [hep-ph]} \BibitemShut {NoStop}%
\bibitem [{\citenamefont {Von~Harling}\ \emph {et~al.}(2020)\citenamefont
  {Von~Harling}, \citenamefont {Pujolas}, and\ \citenamefont
  {Rompineve}}]{vonHarling:2019gme}%
  \BibitemOpen
  \bibfield  {author} {\bibinfo {author} {\bibfnamefont {A.}~\bibnamefont
  {Von~Harling}, \bibfnamefont {B.and~Pomarol}}, \bibinfo {author}
  {\bibfnamefont {O.}~\bibnamefont {Pujolas}},  and\ \bibinfo {author}
  {\bibfnamefont {F.}~\bibnamefont {Rompineve}},\ }\bibfield  {title} {\enquote
  {\bibinfo {title} {\emph{Peccei-Quinn Phase Transition at LIGO}},}\ }\href
  {\doibase 10.1007/JHEP04(2020)195} {\bibfield  {journal} {\bibinfo  {journal}
  {JHEP}\ }\textbf {\bibinfo {volume} {04}},\ \bibinfo {pages} {195} (\bibinfo
  {year} {2020})},\ \Eprint {http://arxiv.org/abs/1912.07587} {arXiv:1912.07587
  [hep-ph]} \BibitemShut {NoStop}%
\bibitem [{\citenamefont {Hall}\ \emph {et~al.}(2020)\citenamefont {Hall},
  \citenamefont {Konstandin}, \citenamefont {McGehee}, \citenamefont
  {Murayama}, and\ \citenamefont {Servant}}]{Hall:2019ank}%
  \BibitemOpen
  \bibfield  {author} {\bibinfo {author} {\bibfnamefont {E.}~\bibnamefont
  {Hall}}, \bibinfo {author} {\bibfnamefont {T.}~\bibnamefont {Konstandin}},
  \bibinfo {author} {\bibfnamefont {R.}~\bibnamefont {McGehee}}, \bibinfo
  {author} {\bibfnamefont {H.}~\bibnamefont {Murayama}},  and\ \bibinfo
  {author} {\bibfnamefont {G.}~\bibnamefont {Servant}},\ }\bibfield  {title}
  {\enquote {\bibinfo {title} {\emph{Baryogenesis From a Dark First-Order Phase
  Transition}},}\ }\href {\doibase 10.1007/JHEP04(2020)042} {\bibfield
  {journal} {\bibinfo  {journal} {JHEP}\ }\textbf {\bibinfo {volume} {04}},\
  \bibinfo {pages} {042} (\bibinfo {year} {2020})},\ \Eprint
  {http://arxiv.org/abs/1910.08068} {arXiv:1910.08068 [hep-ph]} \BibitemShut
  {NoStop}%
\bibitem [{\citenamefont {Huang}\ \emph {et~al.}(2020)\citenamefont {Huang},
  \citenamefont {Sannino}, and\ \citenamefont {Wang}}]{Huang:2020bbe}%
  \BibitemOpen
  \bibfield  {author} {\bibinfo {author} {\bibfnamefont {W.-C.}\ \bibnamefont
  {Huang}}, \bibinfo {author} {\bibfnamefont {F.}~\bibnamefont {Sannino}}, 
  and\ \bibinfo {author} {\bibfnamefont {Z.-W.}\ \bibnamefont {Wang}},\
  }\bibfield  {title} {\enquote {\bibinfo {title} {\emph{Gravitational Waves
  from Pati-Salam Dynamics}},} }\href@noop {} {(\bibinfo {year} {2020})},\
  \Eprint {http://arxiv.org/abs/2004.02332} {arXiv:2004.02332 [hep-ph]}
  \BibitemShut {NoStop}%
\bibitem [{\citenamefont {Fornal}(2020)}]{Fornal:2020ngq}%
  \BibitemOpen
  \bibfield  {author} {\bibinfo {author} {\bibfnamefont {B.}~\bibnamefont
  {Fornal}},\ }\bibfield  {title} {\enquote {\bibinfo {title}
  {\emph{Gravitational Wave Signatures of Lepton Universality Violation}},}
  }\href@noop {} {(\bibinfo {year} {2020})},\ \Eprint
  {http://arxiv.org/abs/2006.08802} {arXiv:2006.08802 [hep-ph]} \BibitemShut
  {NoStop}%
\bibitem [{\citenamefont {Blanco-Pillado} and\ \citenamefont
  {Olum}(2017)}]{Blanco-Pillado:2017oxo}%
  \BibitemOpen
  \bibfield  {author} {\bibinfo {author} {\bibfnamefont {J.~J.}\ \bibnamefont
  {Blanco-Pillado}} and\ \bibinfo {author} {\bibfnamefont {K.~D.}\
  \bibnamefont {Olum}},\ }\bibfield  {title} {\enquote {\bibinfo {title}
  {\emph{Stochastic Gravitational Wave Background from Smoothed Cosmic String
  Loops}},}\ }\href {\doibase 10.1103/PhysRevD.96.104046} {\bibfield  {journal}
  {\bibinfo  {journal} {Phys. Rev. D}\ }\textbf {\bibinfo {volume} {96}},\
  \bibinfo {pages} {104046} (\bibinfo {year} {2017})},\ \Eprint
  {http://arxiv.org/abs/1709.02693} {arXiv:1709.02693 [astro-ph.CO]}
  \BibitemShut {NoStop}%
\bibitem [{\citenamefont {Ringeval} and\ \citenamefont
  {Suyama}(2017)}]{Ringeval:2017eww}%
  \BibitemOpen
  \bibfield  {author} {\bibinfo {author} {\bibfnamefont {C.}~\bibnamefont
  {Ringeval}} and\ \bibinfo {author} {\bibfnamefont {T.}~\bibnamefont
  {Suyama}},\ }\bibfield  {title} {\enquote {\bibinfo {title} {\emph{Stochastic
  Gravitational Waves from Cosmic String Loops in Scaling}},}\ }\href {\doibase
  10.1088/1475-7516/2017/12/027} {\bibfield  {journal} {\bibinfo  {journal}
  {JCAP}\ }\textbf {\bibinfo {volume} {12}},\ \bibinfo {pages} {027} (\bibinfo
  {year} {2017})},\ \Eprint {http://arxiv.org/abs/1709.03845} {arXiv:1709.03845
  [astro-ph.CO]} \BibitemShut {NoStop}%
\bibitem [{\citenamefont {Cui}\ \emph {et~al.}(2018)\citenamefont {Cui},
  \citenamefont {Lewicki}, \citenamefont {Morrissey}, and\ \citenamefont
  {Wells}}]{Cui:2017ufi}%
  \BibitemOpen
  \bibfield  {author} {\bibinfo {author} {\bibfnamefont {Y.}~\bibnamefont
  {Cui}}, \bibinfo {author} {\bibfnamefont {M.}~\bibnamefont {Lewicki}},
  \bibinfo {author} {\bibfnamefont {D.~E.}\ \bibnamefont {Morrissey}},  and\
  \bibinfo {author} {\bibfnamefont {J.~D.}\ \bibnamefont {Wells}},\ }\bibfield
  {title} {\enquote {\bibinfo {title} {\emph{Cosmic Archaeology with
  Gravitational Waves from Cosmic Strings}},}\ }\href {\doibase
  10.1103/PhysRevD.97.123505} {\bibfield  {journal} {\bibinfo  {journal} {Phys.
  Rev. D}\ }\textbf {\bibinfo {volume} {97}},\ \bibinfo {pages} {123505}
  (\bibinfo {year} {2018})},\ \Eprint {http://arxiv.org/abs/1711.03104}
  {arXiv:1711.03104 [hep-ph]} \BibitemShut {NoStop}%
\bibitem [{\citenamefont {Cui}\ \emph {et~al.}(2019)\citenamefont {Cui},
  \citenamefont {Lewicki}, \citenamefont {Morrissey}, and\ \citenamefont
  {Wells}}]{Cui:2018rwi}%
  \BibitemOpen
  \bibfield  {author} {\bibinfo {author} {\bibfnamefont {Y.}~\bibnamefont
  {Cui}}, \bibinfo {author} {\bibfnamefont {M.}~\bibnamefont {Lewicki}},
  \bibinfo {author} {\bibfnamefont {D.~E.}\ \bibnamefont {Morrissey}},  and\
  \bibinfo {author} {\bibfnamefont {J.~D.}\ \bibnamefont {Wells}},\ }\bibfield
  {title} {\enquote {\bibinfo {title} {\emph{Probing the Pre-BBN Universe with
  Gravitational Waves from Cosmic Strings}},}\ }\href {\doibase
  10.1007/JHEP01(2019)081} {\bibfield  {journal} {\bibinfo  {journal} {JHEP}\
  }\textbf {\bibinfo {volume} {01}},\ \bibinfo {pages} {081} (\bibinfo {year}
  {2019})},\ \Eprint {http://arxiv.org/abs/1808.08968} {arXiv:1808.08968
  [hep-ph]} \BibitemShut {NoStop}%
\bibitem [{\citenamefont {Guedes}\ \emph {et~al.}(2018)\citenamefont {Guedes},
  \citenamefont {Avelino}, and\ \citenamefont {Sousa}}]{Guedes:2018afo}%
  \BibitemOpen
  \bibfield  {author} {\bibinfo {author} {\bibfnamefont {G.~S.~F.}\
  \bibnamefont {Guedes}}, \bibinfo {author} {\bibfnamefont {P.~P.}\
  \bibnamefont {Avelino}},  and\ \bibinfo {author} {\bibfnamefont
  {L.}~\bibnamefont {Sousa}},\ }\bibfield  {title} {\enquote {\bibinfo {title}
  {\emph{Signature of Inflation in the Stochastic Gravitational Wave Background
  Generated by Cosmic String Networks}},}\ }\href {\doibase
  10.1103/PhysRevD.98.123505} {\bibfield  {journal} {\bibinfo  {journal} {Phys.
  Rev. D}\ }\textbf {\bibinfo {volume} {98}},\ \bibinfo {pages} {123505}
  (\bibinfo {year} {2018})},\ \Eprint {http://arxiv.org/abs/1809.10802}
  {arXiv:1809.10802 [astro-ph.CO]} \BibitemShut {NoStop}%
\bibitem [{\citenamefont {Dror}\ \emph {et~al.}(2020)\citenamefont {Dror},
  \citenamefont {Hiramatsu}, \citenamefont {Kohri}, \citenamefont {Murayama},\
  and\ \citenamefont {White}}]{Dror:2019syi}%
  \BibitemOpen
  \bibfield  {author} {\bibinfo {author} {\bibfnamefont {J.~A.}\ \bibnamefont
  {Dror}}, \bibinfo {author} {\bibfnamefont {T.}~\bibnamefont {Hiramatsu}},
  \bibinfo {author} {\bibfnamefont {K.}~\bibnamefont {Kohri}}, \bibinfo
  {author} {\bibfnamefont {H.}~\bibnamefont {Murayama}},  and\ \bibinfo
  {author} {\bibfnamefont {G.}~\bibnamefont {White}},\ }\bibfield  {title}
  {\enquote {\bibinfo {title} {\emph{Testing the Seesaw Mechanism and
  Leptogenesis with Gravitational Waves}},}\ }\href {\doibase
  10.1103/PhysRevLett.124.041804} {\bibfield  {journal} {\bibinfo  {journal}
  {Phys. Rev. Lett.}\ }\textbf {\bibinfo {volume} {124}},\ \bibinfo {pages}
  {041804} (\bibinfo {year} {2020})},\ \Eprint
  {http://arxiv.org/abs/1908.03227} {arXiv:1908.03227 [hep-ph]} \BibitemShut
  {NoStop}%
\bibitem [{\citenamefont {Gouttenoire}\ \emph
  {et~al.}(2020{\natexlab{a}})\citenamefont {Gouttenoire}, \citenamefont
  {Servant}, and\ \citenamefont {Simakachorn}}]{Gouttenoire:2019kij}%
  \BibitemOpen
  \bibfield  {author} {\bibinfo {author} {\bibfnamefont {Y.}~\bibnamefont
  {Gouttenoire}}, \bibinfo {author} {\bibfnamefont {G.}~\bibnamefont
  {Servant}},  and\ \bibinfo {author} {\bibfnamefont {P.}~\bibnamefont
  {Simakachorn}},\ }\bibfield  {title} {\enquote {\bibinfo {title}
  {\emph{Beyond the Standard Models with Cosmic Strings}},}\ }\href {\doibase
  10.1088/1475-7516/2020/07/032} {\bibfield  {journal} {\bibinfo  {journal}
  {JCAP}\ }\textbf {\bibinfo {volume} {07}},\ \bibinfo {pages} {032} (\bibinfo
  {year} {2020}{\natexlab{a}})},\ \Eprint {http://arxiv.org/abs/1912.02569}
  {arXiv:1912.02569 [hep-ph]} \BibitemShut {NoStop}%
\bibitem [{\citenamefont {Gouttenoire}\ \emph
  {et~al.}(2020{\natexlab{b}})\citenamefont {Gouttenoire}, \citenamefont
  {Servant}, and\ \citenamefont {Simakachorn}}]{Gouttenoire:2019rtn}%
  \BibitemOpen
  \bibfield  {author} {\bibinfo {author} {\bibfnamefont {Y.}~\bibnamefont
  {Gouttenoire}}, \bibinfo {author} {\bibfnamefont {G.}~\bibnamefont
  {Servant}},  and\ \bibinfo {author} {\bibfnamefont {P.}~\bibnamefont
  {Simakachorn}},\ }\bibfield  {title} {\enquote {\bibinfo {title} {\emph{BSM
  with Cosmic Strings: Heavy, up to EeV Mass, Unstable Particles}},}\ }\href
  {\doibase 10.1088/1475-7516/2020/07/016} {\bibfield  {journal} {\bibinfo
  {journal} {JCAP}\ }\textbf {\bibinfo {volume} {07}},\ \bibinfo {pages} {016}
  (\bibinfo {year} {2020}{\natexlab{b}})},\ \Eprint
  {http://arxiv.org/abs/1912.03245} {arXiv:1912.03245 [hep-ph]} \BibitemShut
  {NoStop}%
\bibitem [{\citenamefont {Buchmuller}\ \emph {et~al.}(2019)\citenamefont
  {Buchmuller}, \citenamefont {Domcke}, \citenamefont {Murayama}, and\
  \citenamefont {Schmitz}}]{Buchmuller:2019gfy}%
  \BibitemOpen
  \bibfield  {author} {\bibinfo {author} {\bibfnamefont {W.}~\bibnamefont
  {Buchmuller}}, \bibinfo {author} {\bibfnamefont {V.}~\bibnamefont {Domcke}},
  \bibinfo {author} {\bibfnamefont {H.}~\bibnamefont {Murayama}},  and\
  \bibinfo {author} {\bibfnamefont {K.}~\bibnamefont {Schmitz}},\ }\bibfield
  {title} {\enquote {\bibinfo {title} {\emph{Probing the Scale of Grand
  Unification with Gravitational Waves}},}\ }\href@noop {} {\  (\bibinfo {year}
  {2019})},\ \Eprint {http://arxiv.org/abs/1912.03695} {arXiv:1912.03695
  [hep-ph]} \BibitemShut {NoStop}%
\bibitem [{\citenamefont {King}\ \emph {et~al.}(2020)\citenamefont {King},
  \citenamefont {Pascoli}, \citenamefont {Turner}, and\ \citenamefont
  {Zhou}}]{King:2020hyd}%
  \BibitemOpen
  \bibfield  {author} {\bibinfo {author} {\bibfnamefont {S.~F.}\ \bibnamefont
  {King}}, \bibinfo {author} {\bibfnamefont {S.}~\bibnamefont {Pascoli}},
  \bibinfo {author} {\bibfnamefont {J.}~\bibnamefont {Turner}},  and\ \bibinfo
  {author} {\bibfnamefont {Y.-L.}\ \bibnamefont {Zhou}},\ }\bibfield  {title}
  {\enquote {\bibinfo {title} {\emph{Gravitational Waves and Proton Decay:
  Complementary Windows into GUTs}},}\ }\href@noop {} {\  (\bibinfo {year}
  {2020})},\ \Eprint {http://arxiv.org/abs/2005.13549} {arXiv:2005.13549
  [hep-ph]} \BibitemShut {NoStop}%
\bibitem [{\citenamefont {Zhou} and\ \citenamefont
  {Bian}(2020)}]{Zhou:2020ils}%
  \BibitemOpen
  \bibfield  {author} {\bibinfo {author} {\bibfnamefont {R.}~\bibnamefont
  {Zhou}} and\ \bibinfo {author} {\bibfnamefont {Ligong}\ \bibnamefont
  {Bian}},\ }\bibfield  {title} {\enquote {\bibinfo {title}
  {\emph{Gravitational Waves from Cosmic Strings and First-Order Phase
  Transition}},}\ }\href@noop {} {\  (\bibinfo {year} {2020})},\ \Eprint
  {http://arxiv.org/abs/2006.13872} {arXiv:2006.13872 [hep-ph]} \BibitemShut
  {NoStop}%
\bibitem [{\citenamefont {Duerr} and\ \citenamefont
  {Fileviez~Perez}(2013)}]{Duerr:2013dza}%
  \BibitemOpen
  \bibfield  {author} {\bibinfo {author} {\bibfnamefont {M.}~\bibnamefont
  {Duerr}}, \bibinfo {author} {\bibfnamefont {P.}\ \bibnamefont
  {Fileviez~Perez}}, and \ \bibinfo {author} {\bibfnamefont {M.~B.}\
  \bibnamefont {Wise}},\ }\bibfield  {title} {\enquote
  {\bibinfo {title} {\emph{Gauge Theory for Baryon and Lepton Numbers with
  Leptoquarks}},}\ }\href {\doibase 10.1103/PhysRevLett.110.231801} {\bibfield
  {journal} {\bibinfo  {journal} {Phys. Rev. Lett.}\ }\textbf {\bibinfo
  {volume} {110}},\ \bibinfo {pages} {231801} (\bibinfo {year} {2013})},\
  \Eprint {http://arxiv.org/abs/1304.0576} {arXiv:1304.0576 [hep-ph]}
  \BibitemShut {NoStop}%
\bibitem [{\citenamefont {Banks} and\ \citenamefont
  {Seiberg}(2011)}]{Banks:2010zn}%
  \BibitemOpen
  \bibfield  {author} {\bibinfo {author} {\bibfnamefont {T.}~\bibnamefont
  {Banks}} and\ \bibinfo {author} {\bibfnamefont {N.}~\bibnamefont
  {Seiberg}},\ }\bibfield  {title} {\enquote {\bibinfo {title}
  {\emph{Symmetries and Strings in Field Theory and Gravity}},}\ }\href
  {\doibase 10.1103/PhysRevD.83.084019} {\bibfield  {journal} {\bibinfo
  {journal} {Phys. Rev. D}\ }\textbf {\bibinfo {volume} {83}},\ \bibinfo
  {pages} {084019} (\bibinfo {year} {2011})},\ \Eprint
  {http://arxiv.org/abs/1011.5120} {arXiv:1011.5120 [hep-th]} \BibitemShut
  {NoStop}%
\bibitem [{\citenamefont {Pais}(1973)}]{Pais:1973mi}%
  \BibitemOpen
  \bibfield  {author} {\bibinfo {author} {\bibfnamefont {A.}~\bibnamefont
  {Pais}},\ }\bibfield  {title} {\enquote {\bibinfo {title} {\emph{Remark on
  Baryon Conservation}},}\ }\href {\doibase 10.1103/PhysRevD.8.1844} {\bibfield
   {journal} {\bibinfo  {journal} {Phys. Rev. D}\ }\textbf {\bibinfo {volume}
  {8}},\ \bibinfo {pages} {1844--1846} (\bibinfo {year} {1973})}\BibitemShut
  {NoStop}%
\bibitem [{\citenamefont {Rajpoot}(1988)}]{Rajpoot:1987yg}%
  \BibitemOpen
  \bibfield  {author} {\bibinfo {author} {\bibfnamefont {S.}~\bibnamefont
  {Rajpoot}},\ }\bibfield  {title} {\enquote {\bibinfo {title} {\emph{Gauge
  Symmetries of Electroweak Interactions}},}\ }\href {\doibase
  10.1007/BF00669312} {\bibfield  {journal} {\bibinfo  {journal} {Int. J.
  Theor. Phys.}\ }\textbf {\bibinfo {volume} {27}},\ \bibinfo {pages} {689}
  (\bibinfo {year} {1988})}\BibitemShut {NoStop}%
\bibitem [{\citenamefont {Foot}\ \emph {et~al.}(1989)\citenamefont {Foot},
  \citenamefont {Joshi}, and\ \citenamefont {Lew}}]{Foot:1989ts}%
  \BibitemOpen
  \bibfield  {author} {\bibinfo {author} {\bibfnamefont {R.}~\bibnamefont
  {Foot}}, \bibinfo {author} {\bibfnamefont {G.~C.}\ \bibnamefont {Joshi}}, \
  and\ \bibinfo {author} {\bibfnamefont {H.}~\bibnamefont {Lew}},\ }\bibfield
  {title} {\enquote {\bibinfo {title} {\emph{Gauged Baryon and Lepton
  Numbers}},}\ }\href {\doibase 10.1103/PhysRevD.40.2487} {\bibfield  {journal}
  {\bibinfo  {journal} {Phys. Rev. D}\ }\textbf {\bibinfo {volume} {40}},\
  \bibinfo {pages} {2487--2489} (\bibinfo {year} {1989})}\BibitemShut {NoStop}%
\bibitem [{\citenamefont {Carone} and\ \citenamefont
  {Murayama}(1995)}]{Carone:1995pu}%
  \BibitemOpen
  \bibfield  {author} {\bibinfo {author} {\bibfnamefont {C.~D.}\ \bibnamefont
  {Carone}} and\ \bibinfo {author} {\bibfnamefont {H.}~\bibnamefont
  {Murayama}},\ }\bibfield  {title} {\enquote {\bibinfo {title}
  {\emph{Realistic Models with a Light U(1) Gauge Boson Coupled to Baryon
  Number}},}\ }\href {\doibase 10.1103/PhysRevD.52.484} {\bibfield  {journal}
  {\bibinfo  {journal} {Phys. Rev. D}\ }\textbf {\bibinfo {volume} {52}},\
  \bibinfo {pages} {484--493} (\bibinfo {year} {1995})},\ \Eprint
  {http://arxiv.org/abs/hep-ph/9501220} {arXiv:hep-ph/9501220} \BibitemShut
  {NoStop}%
\bibitem [{\citenamefont {Georgi} and\ \citenamefont
  {Glashow}(1996)}]{Georgi:1996ei}%
  \BibitemOpen
  \bibfield  {author} {\bibinfo {author} {\bibfnamefont {H.}~\bibnamefont
  {Georgi}} and\ \bibinfo {author} {\bibfnamefont {S.~L.}\ \bibnamefont
  {Glashow}},\ }\bibfield  {title} {\enquote {\bibinfo {title} {\emph{Decays of
  a Leptophobic Gauge Boson}},}\ }\href {\doibase 10.1016/0370-2693(96)00997-5}
  {\bibfield  {journal} {\bibinfo  {journal} {Phys. Lett. B}\ }\textbf
  {\bibinfo {volume} {387}},\ \bibinfo {pages} {341--345} (\bibinfo {year}
  {1996})},\ \Eprint {http://arxiv.org/abs/hep-ph/9607202}
  {arXiv:hep-ph/9607202} \BibitemShut {NoStop}%
\bibitem [{\citenamefont {Fileviez~Perez} and\ \citenamefont
  {Wise}(2010)}]{FileviezPerez:2010gw}%
  \BibitemOpen
  \bibfield  {author} {\bibinfo {author} {\bibfnamefont {P.}~\bibnamefont
  {Fileviez~Perez}} and\ \bibinfo {author} {\bibfnamefont {M.~B.}\
  \bibnamefont {Wise}},\ }\bibfield  {title} {\enquote {\bibinfo {title}
  {\emph{Baryon and Lepton Number as Local Gauge Symmetries}},}\ }\href
  {\doibase 10.1103/PhysRevD.82.079901} {\bibfield  {journal} {\bibinfo
  {journal} {Phys. Rev. D}\ }\textbf {\bibinfo {volume} {82}},\ \bibinfo
  {pages} {011901} (\bibinfo {year} {2010})},\ \bibinfo {note} {[Erratum:
  Phys.Rev.D 82, 079901 (2010)]},\ \Eprint {http://arxiv.org/abs/1002.1754}
  {arXiv:1002.1754 [hep-ph]} \BibitemShut {NoStop}%
\bibitem [{\citenamefont {Schwaller}\ \emph {et~al.}(2013)\citenamefont
  {Schwaller}, \citenamefont {Tait}, and\ \citenamefont
  {Vega-Morales}}]{Schwaller:2013hqa}%
  \BibitemOpen
  \bibfield  {author} {\bibinfo {author} {\bibfnamefont {P.}~\bibnamefont
  {Schwaller}}, \bibinfo {author} {\bibfnamefont {T.~M.~P.}\ \bibnamefont
  {Tait}},  and\ \bibinfo {author} {\bibfnamefont {R.}~\bibnamefont
  {Vega-Morales}},\ }\bibfield  {title} {\enquote {\bibinfo {title} {\emph{Dark
  Matter and Vectorlike Leptons from Gauged Lepton Number}},}\ }\href {\doibase
  10.1103/PhysRevD.88.035001} {\bibfield  {journal} {\bibinfo  {journal} {Phys.
  Rev. D}\ }\textbf {\bibinfo {volume} {88}},\ \bibinfo {pages} {035001}
  (\bibinfo {year} {2013})},\ \Eprint {http://arxiv.org/abs/1305.1108}
  {arXiv:1305.1108 [hep-ph]} \BibitemShut {NoStop}%
\bibitem [{\citenamefont {Fileviez~Perez}\ \emph {et~al.}(2014)\citenamefont
  {Fileviez~Perez}, \citenamefont {Ohmer}, and\ \citenamefont
  {Patel}}]{Perez:2014qfa}%
  \BibitemOpen
  \bibfield  {author} {\bibinfo {author} {\bibfnamefont {P.}~\bibnamefont
  {Fileviez~Perez}}, \bibinfo {author} {\bibfnamefont {S.}~\bibnamefont
  {Ohmer}},  and\ \bibinfo {author} {\bibfnamefont {H.~H.}\ \bibnamefont
  {Patel}},\ }\bibfield  {title} {\enquote {\bibinfo {title} {\emph{Minimal
  Theory for Lepto-Baryons}},}\ }\href {\doibase
  10.1016/j.physletb.2014.06.057} {\bibfield  {journal} {\bibinfo  {journal}
  {Phys. Lett. B}\ }\textbf {\bibinfo {volume} {735}},\ \bibinfo {pages}
  {283--287} (\bibinfo {year} {2014})},\ \Eprint
  {http://arxiv.org/abs/1403.8029} {arXiv:1403.8029 [hep-ph]} \BibitemShut
  {NoStop}%
\bibitem [{\citenamefont {Duerr} and\ \citenamefont
  {Fileviez~Perez}(2015)}]{Duerr:2014wra}%
  \BibitemOpen
  \bibfield  {author} {\bibinfo {author} {\bibfnamefont {M.}~\bibnamefont
  {Duerr}} and\ \bibinfo {author} {\bibfnamefont {P.}~\bibnamefont
  {Fileviez~Perez}},\ }\bibfield  {title} {\enquote {\bibinfo {title}
  {\emph{Theory for Baryon Number and Dark Matter at the LHC}},}\ }\href
  {\doibase 10.1103/PhysRevD.91.095001} {\bibfield  {journal} {\bibinfo
  {journal} {Phys. Rev. D}\ }\textbf {\bibinfo {volume} {91}},\ \bibinfo
  {pages} {095001} (\bibinfo {year} {2015})},\ \Eprint
  {http://arxiv.org/abs/1409.8165} {arXiv:1409.8165 [hep-ph]} \BibitemShut
  {NoStop}%
\bibitem [{\citenamefont {Arnold}\ \emph {et~al.}(2013)\citenamefont {Arnold},
  \citenamefont {Fileviez~Perez}, \citenamefont {Fornal}, and\ \citenamefont
  {Spinner}}]{Arnold:2013qja}%
  \BibitemOpen
  \bibfield  {author} {\bibinfo {author} {\bibfnamefont {J.~M.}\ \bibnamefont
  {Arnold}}, \bibinfo {author} {\bibfnamefont {P.}~\bibnamefont
  {Fileviez~Perez}}, \bibinfo {author} {\bibfnamefont {B.}~\bibnamefont
  {Fornal}},  and\ \bibinfo {author} {\bibfnamefont {S.}~\bibnamefont
  {Spinner}},\ }\bibfield  {title} {\enquote {\bibinfo {title} {\emph{B and L
  at the Supersymmetry Scale, Dark Matter, and R-Parity Violation}},}\ }\href
  {\doibase 10.1103/PhysRevD.88.115009} {\bibfield  {journal} {\bibinfo
  {journal} {Phys. Rev. D}\ }\textbf {\bibinfo {volume} {88}},\ \bibinfo
  {pages} {115009} (\bibinfo {year} {2013})},\ \Eprint
  {http://arxiv.org/abs/1310.7052} {arXiv:1310.7052 [hep-ph]} \BibitemShut
  {NoStop}%
\bibitem [{\citenamefont {Fornal}\ \emph {et~al.}(2015)\citenamefont {Fornal},
  \citenamefont {Rajaraman}, and\ \citenamefont {Tait}}]{Fornal:2015boa}%
  \BibitemOpen
  \bibfield  {author} {\bibinfo {author} {\bibfnamefont {B.}~\bibnamefont
  {Fornal}}, \bibinfo {author} {\bibfnamefont {A.}~\bibnamefont {Rajaraman}}, 
  and\ \bibinfo {author} {\bibfnamefont {T.~M.~P.}\ \bibnamefont {Tait}},\
  }\bibfield  {title} {\enquote {\bibinfo {title} {\emph{Baryon Number as the
  Fourth Color}},}\ }\href {\doibase 10.1103/PhysRevD.92.055022} {\bibfield
  {journal} {\bibinfo  {journal} {Phys. Rev. D}\ }\textbf {\bibinfo {volume}
  {92}},\ \bibinfo {pages} {055022} (\bibinfo {year} {2015})},\ \Eprint
  {http://arxiv.org/abs/1506.06131} {arXiv:1506.06131 [hep-ph]} \BibitemShut
  {NoStop}%
\bibitem [{\citenamefont {Fornal} and\ \citenamefont
  {Tait}(2016)}]{Fornal:2015one}%
  \BibitemOpen
  \bibfield  {author} {\bibinfo {author} {\bibfnamefont {B.}~\bibnamefont
  {Fornal}} and\ \bibinfo {author} {\bibfnamefont {T.~M.~P.}\ \bibnamefont
  {Tait}},\ }\bibfield  {title} {\enquote {\bibinfo {title} {\emph{Dark Matter
  from Unification of Color and Baryon Number}},}\ }\href {\doibase
  10.1103/PhysRevD.93.075010} {\bibfield  {journal} {\bibinfo  {journal} {Phys.
  Rev. D}\ }\textbf {\bibinfo {volume} {93}},\ \bibinfo {pages} {075010}
  (\bibinfo {year} {2016})},\ \Eprint {http://arxiv.org/abs/1511.07380}
  {arXiv:1511.07380 [hep-ph]} \BibitemShut {NoStop}%
\bibitem [{\citenamefont {Fileviez~Perez} and\ \citenamefont
  {Ohmer}(2017)}]{FileviezPerez:2016laj}%
  \BibitemOpen
  \bibfield  {author} {\bibinfo {author} {\bibfnamefont {P.}~\bibnamefont
  {Fileviez~Perez}} and\ \bibinfo {author} {\bibfnamefont {S.}~\bibnamefont
  {Ohmer}},\ }\bibfield  {title} {\enquote {\bibinfo {title} {\emph{Unification
  and Local Baryon Number}},}\ }\href {\doibase 10.1016/j.physletb.2017.02.049}
  {\bibfield  {journal} {\bibinfo  {journal} {Phys. Lett. B}\ }\textbf
  {\bibinfo {volume} {768}},\ \bibinfo {pages} {86--91} (\bibinfo {year}
  {2017})},\ \Eprint {http://arxiv.org/abs/1612.07165} {arXiv:1612.07165
  [hep-ph]} \BibitemShut {NoStop}%
\bibitem [{\citenamefont {Fornal}\ \emph {et~al.}(2017)\citenamefont {Fornal},
  \citenamefont {Shirman}, \citenamefont {Tait}, and\ \citenamefont
  {West}}]{Fornal:2017owa}%
  \BibitemOpen
  \bibfield  {author} {\bibinfo {author} {\bibfnamefont {B.}~\bibnamefont
  {Fornal}}, \bibinfo {author} {\bibfnamefont {Y.}~\bibnamefont {Shirman}},
  \bibinfo {author} {\bibfnamefont {T.~M.~P.}\ \bibnamefont {Tait}},  and\
  \bibinfo {author} {\bibfnamefont {J.~R.}\ \bibnamefont {West}},\ }\bibfield
  {title} {\enquote {\bibinfo {title} {\emph{Asymmetric Dark Matter and
  Baryogenesis from $SU(2)_{\ell}$}},}\ }\href {\doibase
  10.1103/PhysRevD.96.035001} {\bibfield  {journal} {\bibinfo  {journal} {Phys.
  Rev. D}\ }\textbf {\bibinfo {volume} {96}},\ \bibinfo {pages} {035001}
  (\bibinfo {year} {2017})},\ \Eprint {http://arxiv.org/abs/1703.00199}
  {arXiv:1703.00199 [hep-ph]} \BibitemShut {NoStop}%
\bibitem [{\citenamefont {Aghanim}\ \emph {et~al.}(2018)\citenamefont {Aghanim}
  \emph {et~al.}}]{Aghanim:2018eyx}%
  \BibitemOpen
  \bibfield  {author} {\bibinfo {author} {\bibfnamefont {N.}~\bibnamefont
  {Aghanim}} \emph {et~al.} (\bibinfo {collaboration} {Planck}),\ }\bibfield
  {title} {\enquote {\bibinfo {title} {\emph{Planck 2018 Results. VI.
  Cosmological Parameters}},}\ }\href@noop {} {\  (\bibinfo {year} {2018})},\
  \Eprint {http://arxiv.org/abs/1807.06209} {arXiv:1807.06209 [astro-ph.CO]}
  \BibitemShut {NoStop}%
\bibitem [{\citenamefont {Kibble}(1976)}]{Kibble:1976sj}%
  \BibitemOpen
  \bibfield  {author} {\bibinfo {author} {\bibfnamefont {T.~W.~B.}\
  \bibnamefont {Kibble}},\ }\bibfield  {title} {\enquote {\bibinfo {title}
  {\emph{Topology of Cosmic Domains and Strings}},}\ }\href {\doibase
  10.1088/0305-4470/9/8/029} {\bibfield  {journal} {\bibinfo  {journal} {J.
  Phys. A}\ }\textbf {\bibinfo {volume} {9}},\ \bibinfo {pages} {1387--1398}
  (\bibinfo {year} {1976})}\BibitemShut {NoStop}%
\bibitem [{\citenamefont {Vilenkin} and\ \citenamefont
  {Shellard}(2000)}]{Vilenkin:2000jqa}%
  \BibitemOpen
  \bibfield  {author} {\bibinfo {author} {\bibfnamefont {A.}~\bibnamefont
  {Vilenkin}} and\ \bibinfo {author} {\bibfnamefont {E.~P.~S.}\ \bibnamefont
  {Shellard}},\ }\bibfield  {title} {\enquote {\bibinfo {title} {\emph{Cosmic
  Strings and Other Topological Defects}},}\ }\href@noop {} {\bibfield
  {journal} {\bibinfo  {journal} {Cambridge University Press}\ } (\bibinfo
  {year} {2000})}\BibitemShut {NoStop}%
\bibitem [{\citenamefont {Ade}\ \emph {et~al.}(2014)\citenamefont {Ade} \emph
  {et~al.}}]{Ade:2013xla}%
  \BibitemOpen
  \bibfield  {author} {\bibinfo {author} {\bibfnamefont {P.~A.~R.}\
  \bibnamefont {Ade}} \emph {et~al.} (\bibinfo {collaboration} {Planck}),\
  }\bibfield  {title} {\enquote {\bibinfo {title} {\emph{Planck 2013 Results.
  XXV. Searches for Cosmic Strings and Other Topological Defects}},}\ }\href
  {\doibase 10.1051/0004-6361/201321621} {\bibfield  {journal} {\bibinfo
  {journal} {Astron. Astrophys.}\ }\textbf {\bibinfo {volume} {571}},\ \bibinfo
  {pages} {A25} (\bibinfo {year} {2014})},\ \Eprint
  {http://arxiv.org/abs/1303.5085} {arXiv:1303.5085 [astro-ph.CO]} \BibitemShut
  {NoStop}%
\bibitem [{\citenamefont {Kibble}(1985)}]{Kibble:1984hp}%
  \BibitemOpen
  \bibfield  {author} {\bibinfo {author} {\bibfnamefont {T.~W.~B.}\
  \bibnamefont {Kibble}},\ }\bibfield  {title} {\enquote {\bibinfo {title}
  {\emph{Evolution of a System of Cosmic Strings}},}\ }\href {\doibase
  10.1016/0550-3213(85)90596-6} {\bibfield  {journal} {\bibinfo  {journal}
  {Nucl. Phys. B}\ }\textbf {\bibinfo {volume} {252}},\ \bibinfo {pages} {227}
  (\bibinfo {year} {1985})},\ \bibinfo {note} {[Erratum: Nucl.Phys.B 261, 750
  (1985)]}\vspace{1mm}\BibitemShut {NoStop}%
\bibitem [{\citenamefont {Bennett} and\ \citenamefont
  {Bouchet}(1988)}]{PhysRevLett.60.257}%
  \BibitemOpen
  \bibfield  {author} {\bibinfo {author} {\bibfnamefont {D.~P.}\ \bibnamefont
  {Bennett}} and\ \bibinfo {author} {\bibfnamefont {F.~R.}\ \bibnamefont
  {Bouchet}},\ }\bibfield  {title} {\enquote {\bibinfo {title} {\emph{Evidence
  for a Scaling Solution in Cosmic-String Evolution}},}\ }\href {\doibase
  10.1103/PhysRevLett.60.257} {\bibfield  {journal} {\bibinfo  {journal} {Phys.
  Rev. Lett.}\ }\textbf {\bibinfo {volume} {60}},\ \bibinfo {pages} {257--260}
  (\bibinfo {year} {1988})}\BibitemShut {NoStop}%
\bibitem [{\citenamefont {Bennett} and\ \citenamefont
  {Bouchet}(1989)}]{PhysRevLett.63.2776}%
  \BibitemOpen
  \bibfield  {author} {\bibinfo {author} {\bibfnamefont {D.~P.}\ \bibnamefont
  {Bennett}} and\ \bibinfo {author} {\bibfnamefont {F.~R.}\ \bibnamefont
  {Bouchet}},\ }\bibfield  {title} {\enquote {\bibinfo {title}
  {\emph{Cosmic-String Evolution}},}\ }\href {\doibase
  10.1103/PhysRevLett.63.2776} {\bibfield  {journal} {\bibinfo  {journal}
  {Phys. Rev. Lett.}\ }\textbf {\bibinfo {volume} {63}},\ \bibinfo {pages}
  {2776--2779} (\bibinfo {year} {1989})}\BibitemShut {NoStop}%
\bibitem [{\citenamefont {Albrecht} and\ \citenamefont
  {Turok}(1989)}]{PhysRevD.40.973}%
  \BibitemOpen
  \bibfield  {author} {\bibinfo {author} {\bibfnamefont {A.}~\bibnamefont
  {Albrecht}} and\ \bibinfo {author} {\bibfnamefont {N.}~\bibnamefont
  {Turok}},\ }\bibfield  {title} {\enquote {\bibinfo {title} {\emph{Evolution
  of Cosmic String Networks}},}\ }\href {\doibase 10.1103/PhysRevD.40.973}
  {\bibfield  {journal} {\bibinfo  {journal} {Phys. Rev. D}\ }\textbf {\bibinfo
  {volume} {40}},\ \bibinfo {pages} {973--1001} (\bibinfo {year}
  {1989})}\BibitemShut {NoStop}%
\bibitem [{\citenamefont {Allen} and\ \citenamefont
  {Shellard}(1990)}]{PhysRevLett.64.119}%
  \BibitemOpen
  \bibfield  {author} {\bibinfo {author} {\bibfnamefont {B.}~\bibnamefont
  {Allen}} and\ \bibinfo {author} {\bibfnamefont {E.~P.~S.}\ \bibnamefont
  {Shellard}},\ }\bibfield  {title} {\enquote {\bibinfo {title}
  {\emph{Cosmic-String Evolution: A Numerical Simulation}},}\ }\href {\doibase
  10.1103/PhysRevLett.64.119} {\bibfield  {journal} {\bibinfo  {journal} {Phys.
  Rev. Lett.}\ }\textbf {\bibinfo {volume} {64}},\ \bibinfo {pages} {119--122}
  (\bibinfo {year} {1990})}\BibitemShut {NoStop}%
\bibitem [{\citenamefont {Hindmarsh} and\ \citenamefont
  {Kibble}(1995)}]{Hindmarsh:1994re}%
  \BibitemOpen
  \bibfield  {author} {\bibinfo {author} {\bibfnamefont {M.~B.}\ \bibnamefont
  {Hindmarsh}} and\ \bibinfo {author} {\bibfnamefont {T.~W.~B.}\ \bibnamefont
  {Kibble}},\ }\bibfield  {title} {\enquote {\bibinfo {title} {\emph{Cosmic
  Strings}},}\ }\href {\doibase 10.1088/0034-4885/58/5/001} {\bibfield
  {journal} {\bibinfo  {journal} {Rept. Prog. Phys.}\ }\textbf {\bibinfo
  {volume} {58}},\ \bibinfo {pages} {477--562} (\bibinfo {year} {1995})},\
  \Eprint {http://arxiv.org/abs/hep-ph/9411342} {arXiv:hep-ph/9411342}
  \BibitemShut {NoStop}%
\bibitem [{\citenamefont {Olum} and\ \citenamefont
  {Blanco-Pillado}(2000)}]{Olum:1999sg}%
  \BibitemOpen
  \bibfield  {author} {\bibinfo {author} {\bibfnamefont {K.~D.}\ \bibnamefont
  {Olum}} and\ \bibinfo {author} {\bibfnamefont {J.~J.}\ \bibnamefont
  {Blanco-Pillado}},\ }\bibfield  {title} {\enquote {\bibinfo {title}
  {\emph{Radiation from Cosmic String Standing Waves}},}\ }\href {\doibase
  10.1103/PhysRevLett.84.4288} {\bibfield  {journal} {\bibinfo  {journal}
  {Phys. Rev. Lett.}\ }\textbf {\bibinfo {volume} {84}},\ \bibinfo {pages}
  {4288--4291} (\bibinfo {year} {2000})},\ \Eprint
  {http://arxiv.org/abs/astro-ph/9910354} {arXiv:astro-ph/9910354} \BibitemShut
  {NoStop}%
\bibitem [{\citenamefont {Moore}\ \emph {et~al.}(2002)\citenamefont {Moore},
  \citenamefont {Shellard}, and\ \citenamefont {Martins}}]{Moore:2001px}%
  \BibitemOpen
  \bibfield  {author} {\bibinfo {author} {\bibfnamefont {J.~N.}\ \bibnamefont
  {Moore}}, \bibinfo {author} {\bibfnamefont {E.~P.~S.}\ \bibnamefont
  {Shellard}},  and\ \bibinfo {author} {\bibfnamefont {C.~J.~A.~P.}\
  \bibnamefont {Martins}},\ }\bibfield  {title} {\enquote {\bibinfo {title}
  {\emph{On the Evolution of Abelian-Higgs String Networks}},}\ }\href
  {\doibase 10.1103/PhysRevD.65.023503} {\bibfield  {journal} {\bibinfo
  {journal} {Phys. Rev. D}\ }\textbf {\bibinfo {volume} {65}},\ \bibinfo
  {pages} {023503} (\bibinfo {year} {2002})},\ \Eprint
  {http://arxiv.org/abs/hep-ph/0107171} {arXiv:hep-ph/0107171} \BibitemShut
  {NoStop}%
\bibitem [{\citenamefont {Blanco-Pillado}\ \emph {et~al.}(2014)\citenamefont
  {Blanco-Pillado}, \citenamefont {Olum}, and\ \citenamefont
  {Shlaer}}]{Blanco-Pillado:2013qja}%
  \BibitemOpen
  \bibfield  {author} {\bibinfo {author} {\bibfnamefont {J.~J.}\ \bibnamefont
  {Blanco-Pillado}}, \bibinfo {author} {\bibfnamefont {K.~D.}\ \bibnamefont
  {Olum}},  and\ \bibinfo {author} {\bibfnamefont {B.}~\bibnamefont
  {Shlaer}},\ }\bibfield  {title} {\enquote {\bibinfo {title} {\emph{The Number
  of Cosmic String Loops}},}\ }\href {\doibase 10.1103/PhysRevD.89.023512}
  {\bibfield  {journal} {\bibinfo  {journal} {Phys. Rev. D}\ }\textbf {\bibinfo
  {volume} {89}},\ \bibinfo {pages} {023512} (\bibinfo {year} {2014})},\
  \Eprint {http://arxiv.org/abs/1309.6637} {arXiv:1309.6637 [astro-ph.CO]}
  \BibitemShut {NoStop}%
\bibitem [{\citenamefont {Vachaspati} and\ \citenamefont
  {Vilenkin}(1985)}]{PhysRevD.31.3052}%
  \BibitemOpen
  \bibfield  {author} {\bibinfo {author} {\bibfnamefont {T.}~\bibnamefont
  {Vachaspati}} and\ \bibinfo {author} {\bibfnamefont {A.}~\bibnamefont
  {Vilenkin}},\ }\bibfield  {title} {\enquote {\bibinfo {title}
  {\emph{Gravitational Radiation from Cosmic Strings}},}\ }\href {\doibase
  10.1103/PhysRevD.31.3052} {\bibfield  {journal} {\bibinfo  {journal} {Phys.
  Rev. D}\ }\textbf {\bibinfo {volume} {31}},\ \bibinfo {pages} {3052--3058}
  (\bibinfo {year} {1985})}\BibitemShut {NoStop}%
\bibitem [{\citenamefont {Martins} and\ \citenamefont
  {Shellard}(1996{\natexlab{a}})}]{Martins:1995tg}%
  \BibitemOpen
  \bibfield  {author} {\bibinfo {author} {\bibfnamefont {C.~J.~A.~P.}\
  \bibnamefont {Martins}} and\ \bibinfo {author} {\bibfnamefont {E.~P.~S.}\
  \bibnamefont {Shellard}},\ }\bibfield  {title} {\enquote {\bibinfo {title}
  {\emph{String Evolution with Friction}},}\ }\href {\doibase
  10.1103/PhysRevD.53.R575} {\bibfield  {journal} {\bibinfo  {journal} {Phys.
  Rev. D}\ }\textbf {\bibinfo {volume} {53}},\ \bibinfo {pages} {575--579}
  (\bibinfo {year} {1996}{\natexlab{a}})},\ \Eprint
  {http://arxiv.org/abs/hep-ph/9507335} {arXiv:hep-ph/9507335} \BibitemShut
  {NoStop}%
\bibitem [{\citenamefont {Martins} and\ \citenamefont
  {Shellard}(1996{\natexlab{b}})}]{Martins:1996jp}%
  \BibitemOpen
  \bibfield  {author} {\bibinfo {author} {\bibfnamefont {C.~J.~A.~P.}\
  \bibnamefont {Martins}} and\ \bibinfo {author} {\bibfnamefont {E.~P.~S.}\
  \bibnamefont {Shellard}},\ }\bibfield  {title} {\enquote {\bibinfo {title}
  {\emph{Quantitative String Evolution}},}\ }\href {\doibase
  10.1103/PhysRevD.54.2535} {\bibfield  {journal} {\bibinfo  {journal} {Phys.
  Rev. D}\ }\textbf {\bibinfo {volume} {54}},\ \bibinfo {pages} {2535--2556}
  (\bibinfo {year} {1996}{\natexlab{b}})},\ \Eprint
  {http://arxiv.org/abs/hep-ph/9602271} {arXiv:hep-ph/9602271} \BibitemShut
  {NoStop}%
\bibitem [{\citenamefont {Martins} and\ \citenamefont
  {Shellard}(2002)}]{Martins:2000cs}%
  \BibitemOpen
  \bibfield  {author} {\bibinfo {author} {\bibfnamefont {C.~J.~A.~P.}\
  \bibnamefont {Martins}} and\ \bibinfo {author} {\bibfnamefont {E.~P.~S.}\
  \bibnamefont {Shellard}},\ }\bibfield  {title} {\enquote {\bibinfo {title}
  {\emph{Extending the Velocity Dependent One Scale String Evolution Model}},}\
  }\href {\doibase 10.1103/PhysRevD.65.043514} {\bibfield  {journal} {\bibinfo
  {journal} {Phys. Rev. D}\ }\textbf {\bibinfo {volume} {65}},\ \bibinfo
  {pages} {043514} (\bibinfo {year} {2002})},\ \Eprint
  {http://arxiv.org/abs/hep-ph/0003298} {arXiv:hep-ph/0003298} \BibitemShut
  {NoStop}%
\bibitem [{\citenamefont {Guo}\ \emph {et~al.}(2020)\citenamefont {Guo},
  \citenamefont {Sinha}, \citenamefont {Vagie}, and\ \citenamefont
  {White}}]{Guo:2020grp}%
  \BibitemOpen
  \bibfield  {author} {\bibinfo {author} {\bibfnamefont {H.-K.}\ \bibnamefont
  {Guo}}, \bibinfo {author} {\bibfnamefont {K.}~\bibnamefont {Sinha}}, \bibinfo
  {author} {\bibfnamefont {D.}~\bibnamefont {Vagie}},  and\ \bibinfo {author}
  {\bibfnamefont {G.}~\bibnamefont {White}},\ }\bibfield  {title} {\enquote
  {\bibinfo {title} {\emph{Phase Transitions in an Expanding Universe:
  Stochastic Gravitational Waves in Standard and Non-Standard Histories}},}\
  }\href@noop {} {\  (\bibinfo {year} {2020})},\ \Eprint
  {http://arxiv.org/abs/2007.08537} {arXiv:2007.08537 [hep-ph]} \BibitemShut
  {NoStop}%
\bibitem [{\citenamefont {Linde}(1983)}]{LINDE1983421}%
  \BibitemOpen
  \bibfield  {author} {\bibinfo {author} {\bibfnamefont {A.~D.}\ \bibnamefont
  {Linde}},\ }\bibfield  {title} {\enquote {\bibinfo {title} {\emph{Decay of
  the False Vacuum at Finite Temperature}},}\ }\href {\doibase
  https://doi.org/10.1016/0550-3213(83)90293-6} {\bibfield  {journal} {\bibinfo
   {journal} {Nuclear Physics B}\ }\textbf {\bibinfo {volume} {216}},\ \bibinfo
  {pages} {421 -- 445} (\bibinfo {year} {1983})}\vspace{15mm}\BibitemShut {NoStop}%
\bibitem [{\citenamefont {Espinosa}\ \emph {et~al.}(2010)\citenamefont
  {Espinosa}, \citenamefont {Konstandin}, \citenamefont {No}, and\
  \citenamefont {Servant}}]{Espinosa:2010hh}%
  \BibitemOpen
  \bibfield  {author} {\bibinfo {author} {\bibfnamefont {J.~R.}\ \bibnamefont
  {Espinosa}}, \bibinfo {author} {\bibfnamefont {T.}~\bibnamefont
  {Konstandin}}, \bibinfo {author} {\bibfnamefont {J.~M.}\ \bibnamefont {No}},
   and\ \bibinfo {author} {\bibfnamefont {G.}~\bibnamefont {Servant}},\
  }\bibfield  {title} {\enquote {\bibinfo {title} {\emph{Energy Budget of
  Cosmological First-Order Phase Transitions}},}\ }\href {\doibase
  10.1088/1475-7516/2010/06/028} {\bibfield  {journal} {\bibinfo  {journal}
  {JCAP}\ }\textbf {\bibinfo {volume} {06}},\ \bibinfo {pages} {028} (\bibinfo
  {year} {2010})},\ \Eprint {http://arxiv.org/abs/1004.4187} {arXiv:1004.4187
  [hep-ph]} \BibitemShut {NoStop}%
\bibitem [{\citenamefont {Hindmarsh}\ \emph {et~al.}(2014)\citenamefont
  {Hindmarsh}, \citenamefont {Huber}, \citenamefont {Rummukainen}, and\
  \citenamefont {Weir}}]{Hindmarsh:2013xza}%
  \BibitemOpen
  \bibfield  {author} {\bibinfo {author} {\bibfnamefont {M.}~\bibnamefont
  {Hindmarsh}}, \bibinfo {author} {\bibfnamefont {S.~J.}\ \bibnamefont
  {Huber}}, \bibinfo {author} {\bibfnamefont {K.}~\bibnamefont {Rummukainen}},
   and\ \bibinfo {author} {\bibfnamefont {D.~J.}\ \bibnamefont {Weir}},\
  }\bibfield  {title} {\enquote {\bibinfo {title} {\emph{Gravitational Waves
  from the Sound of a First Order Phase Transition}},}\ }\href {\doibase
  10.1103/PhysRevLett.112.041301} {\bibfield  {journal} {\bibinfo  {journal}
  {Phys. Rev. Lett.}\ }\textbf {\bibinfo {volume} {112}},\ \bibinfo {pages}
  {041301} (\bibinfo {year} {2014})},\ \Eprint {http://arxiv.org/abs/1304.2433}
  {arXiv:1304.2433 [hep-ph]} \BibitemShut {NoStop}%
\bibitem [{\citenamefont {Caprini}\ \emph {et~al.}(2016)\citenamefont {Caprini}
  \emph {et~al.}}]{Caprini:2015zlo}%
  \BibitemOpen
  \bibfield  {author} {\bibinfo {author} {\bibfnamefont {C.}~\bibnamefont
  {Caprini}} \emph {et~al.},\ }\bibfield  {title} {\enquote {\bibinfo {title}
  {\emph{Science with the Space-Based Interferometer eLISA. II: Gravitational
  Waves from Cosmological Phase Transitions}},}\ }\href {\doibase
  10.1088/1475-7516/2016/04/001} {\bibfield  {journal} {\bibinfo  {journal}
  {JCAP}\ }\textbf {\bibinfo {volume} {04}},\ \bibinfo {pages} {001} (\bibinfo
  {year} {2016})},\ \Eprint {http://arxiv.org/abs/1512.06239} {arXiv:1512.06239
  [astro-ph.CO]} \BibitemShut {NoStop}%
\bibitem [{\citenamefont {Hindmarsh}\ \emph {et~al.}(2017)\citenamefont
  {Hindmarsh}, \citenamefont {Huber}, \citenamefont {Rummukainen}, and\
  \citenamefont {Weir}}]{Hindmarsh:2017gnf}%
  \BibitemOpen
  \bibfield  {author} {\bibinfo {author} {\bibfnamefont {M.}~\bibnamefont
  {Hindmarsh}}, \bibinfo {author} {\bibfnamefont {S.~J.}\ \bibnamefont
  {Huber}}, \bibinfo {author} {\bibfnamefont {K.}~\bibnamefont {Rummukainen}},
   and\ \bibinfo {author} {\bibfnamefont {D.~J.}\ \bibnamefont {Weir}},\
  }\bibfield  {title} {\enquote {\bibinfo {title} {\emph{Shape of the Acoustic
  Gravitational Wave Power Spectrum from a First Order Phase Transition}},}\
  }\href {\doibase 10.1103/PhysRevD.96.103520} {\bibfield  {journal} {\bibinfo
  {journal} {Phys. Rev. D}\ }\textbf {\bibinfo {volume} {96}},\ \bibinfo
  {pages} {103520} (\bibinfo {year} {2017})},\ \bibinfo {note} {[Erratum:
  Phys.Rev.D 101, 089902 (2020)]},\ \Eprint {http://arxiv.org/abs/1704.05871}
  {arXiv:1704.05871 [astro-ph.CO]} \BibitemShut {NoStop}%
\bibitem [{\citenamefont {Ellis}\ \emph {et~al.}(2019)\citenamefont {Ellis},
  \citenamefont {Lewicki}, and\ \citenamefont {No}}]{Ellis:2018mja}%
  \BibitemOpen
  \bibfield  {author} {\bibinfo {author} {\bibfnamefont {J.}~\bibnamefont
  {Ellis}}, \bibinfo {author} {\bibfnamefont {M.}~\bibnamefont {Lewicki}}, \
  and\ \bibinfo {author} {\bibfnamefont {J.~M.}\ \bibnamefont {No}},\
  }\bibfield  {title} {\enquote {\bibinfo {title} {\emph{On the Maximal
  Strength of a First-Order Electroweak Phase Transition and its Gravitational
  Wave Signal}},}\ }\href {\doibase 10.1088/1475-7516/2019/04/003} {\bibfield
  {journal} {\bibinfo  {journal} {JCAP}\ }\textbf {\bibinfo {volume} {04}},\
  \bibinfo {pages} {003} (\bibinfo {year} {2019})},\ \Eprint
  {http://arxiv.org/abs/1809.08242} {arXiv:1809.08242 [hep-ph]} \BibitemShut
  {NoStop}%
\bibitem [{\citenamefont {Yagi} and\ \citenamefont
  {Seto}(2011)}]{Yagi:2011wg}%
  \BibitemOpen
  \bibfield  {author} {\bibinfo {author} {\bibfnamefont {K.}~\bibnamefont
  {Yagi}} and\ \bibinfo {author} {\bibfnamefont {N.}~\bibnamefont {Seto}},\
  }\bibfield  {title} {\enquote {\bibinfo {title} {\emph{Detector Configuration
  of DECIGO/BBO and Identification of Cosmological Neutron-Star Binaries}},}\
  }\href {\doibase 10.1103/PhysRevD.83.044011} {\bibfield  {journal} {\bibinfo
  {journal} {Phys. Rev. D}\ }\textbf {\bibinfo {volume} {83}},\ \bibinfo
  {pages} {044011} (\bibinfo {year} {2011})},\ \bibinfo {note} {[Erratum:
  Phys.Rev.D 95, 109901 (2017)]},\ \Eprint {http://arxiv.org/abs/1101.3940}
  {arXiv:1101.3940 [astro-ph.CO]} \BibitemShut {NoStop}%
\bibitem [{\citenamefont {Sathyaprakash}\ \emph {et~al.}(2012)\citenamefont
  {Sathyaprakash} \emph {et~al.}}]{Sathyaprakash:2012jk}%
  \BibitemOpen
  \bibfield  {author} {\bibinfo {author} {\bibfnamefont {B.}~\bibnamefont
  {Sathyaprakash}} \emph {et~al.},\ }\bibfield  {title} {\enquote {\bibinfo
  {title} {\emph{Scientific Objectives of Einstein Telescope}},}\ }\href
  {\doibase 10.1088/0264-9381/29/12/124013} {\bibfield  {journal} {\bibinfo
  {journal} {Class. Quant. Grav.}\ }\textbf {\bibinfo {volume} {29}},\ \bibinfo
  {pages} {124013} (\bibinfo {year} {2012})},\ \bibinfo {note} {[Erratum:
  Class.Quant.Grav. 30, 079501 (2013)]},\ \Eprint
  {http://arxiv.org/abs/1206.0331} {arXiv:1206.0331 [gr-qc]} \BibitemShut
  {NoStop}%
\bibitem [{\citenamefont {Kosowsky}\ \emph {et~al.}(1992)\citenamefont
  {Kosowsky}, \citenamefont {Turner}, and\ \citenamefont
  {Watkins}}]{Kosowsky:1991ua}%
  \BibitemOpen
  \bibfield  {author} {\bibinfo {author} {\bibfnamefont {A.}~\bibnamefont
  {Kosowsky}}, \bibinfo {author} {\bibfnamefont {M.~S.}\ \bibnamefont
  {Turner}},  and\ \bibinfo {author} {\bibfnamefont {R.}~\bibnamefont
  {Watkins}},\ }\bibfield  {title} {\enquote {\bibinfo {title}
  {\emph{Gravitational Radiation from Colliding Vacuum Bubbles}},}\ }\href
  {\doibase 10.1103/PhysRevD.45.4514} {\bibfield  {journal} {\bibinfo
  {journal} {Phys. Rev. D}\ }\textbf {\bibinfo {volume} {45}},\ \bibinfo
  {pages} {4514--4535} (\bibinfo {year} {1992})}\BibitemShut {NoStop}%
\bibitem [{\citenamefont {Huber} and\ \citenamefont
  {Konstandin}(2008)}]{Huber:2008hg}%
  \BibitemOpen
  \bibfield  {author} {\bibinfo {author} {\bibfnamefont {S.~J.}\ \bibnamefont
  {Huber}} and\ \bibinfo {author} {\bibfnamefont {T.}~\bibnamefont
  {Konstandin}},\ }\bibfield  {title} {\enquote {\bibinfo {title}
  {\emph{Gravitational Wave Production by Collisions: More Bubbles}},}\ }\href
  {\doibase 10.1088/1475-7516/2008/09/022} {\bibfield  {journal} {\bibinfo
  {journal} {JCAP}\ }\textbf {\bibinfo {volume} {09}},\ \bibinfo {pages} {022}
  (\bibinfo {year} {2008})},\ \Eprint {http://arxiv.org/abs/0806.1828}
  {arXiv:0806.1828 [hep-ph]} \BibitemShut {NoStop}%
\bibitem [{\citenamefont {Kamionkowski}\ \emph {et~al.}(1994)\citenamefont
  {Kamionkowski}, \citenamefont {Kosowsky}, and\ \citenamefont
  {Turner}}]{Kamionkowski:1993fg}%
  \BibitemOpen
  \bibfield  {author} {\bibinfo {author} {\bibfnamefont {M.}~\bibnamefont
  {Kamionkowski}}, \bibinfo {author} {\bibfnamefont {A.}~\bibnamefont
  {Kosowsky}},  and\ \bibinfo {author} {\bibfnamefont {M.~S.}\ \bibnamefont
  {Turner}},\ }\bibfield  {title} {\enquote {\bibinfo {title}
  {\emph{Gravitational Radiation from First Order Phase Transitions}},}\ }\href
  {\doibase 10.1103/PhysRevD.49.2837} {\bibfield  {journal} {\bibinfo
  {journal} {Phys. Rev. D}\ }\textbf {\bibinfo {volume} {49}},\ \bibinfo
  {pages} {2837--2851} (\bibinfo {year} {1994})},\ \Eprint
  {http://arxiv.org/abs/astro-ph/9310044} {arXiv:astro-ph/9310044} \BibitemShut
  {NoStop}%
\end{thebibliography}%

\end{document}